\documentclass[aps,prl,preprint,superscriptaddress]{revtex4-2}

\usepackage{graphicx}% Include figure files
\usepackage{dcolumn}% Align table columns on decimal point
\usepackage{bm}% bold math
\usepackage{color}

\begin{document}
	
	\title{Non-equilibrium carrier dynamics and band structure of graphene on 2D tin}
	
	% repeat the \author .. \affiliation  etc. as needed
	% \email, \thanks, \homepage, \altaffiliation all apply to the current
	% author. Explanatory text should go in the []'s, actual e-mail
	% address or url should go in the {}'s for \email and \homepage.
	% Please use the appropriate macro foreach each type of information
	
	% \affiliation command applies to all authors since the last
	% \affiliation command. The \affiliation command should follow the
	% other information
	% \affiliation can be followed by \email, \homepage, \thanks as well.
	
	\author{Maria-Elisabeth Federl}
	\affiliation{Department for Experimental and Applied Physics, University of Regensburg, 93040 Regensburg, Germany}
	
	\author{Niklas Witt}
	\affiliation{I. Institute of Theoretical Physics, University of Hamburg, 22607 Hamburg, Germany}
	\affiliation{The Hamburg Centre for Ultrafast Imaging, 22761 Hamburg, Germany}
	
	\author{Biao Yang}
	\affiliation{Physics Department E20, TUM School of Natural Sciences, Technical University of Munich, 85748 Garching, Germany}
	
	\author{Leonard Weigl}
	\author{Niklas Hofmann}
	\author{Johannes Gradl}
	\affiliation{Department for Experimental and Applied Physics, University of Regensburg, 93040 Regensburg, Germany}
	
	\author{Ignacio Piquero-Zulaica}
	\author{Johannes V. Barth}
	\affiliation{Physics Department E20, TUM School of Natural Sciences, Technical University of Munich, 85748 Garching, Germany}
	
	\author{Neeraj Mishra}
	\affiliation{Center for Nanotechnology Innovation@NEST, Istituto Italiano di Tecnologia, Pisa, Italy}
	\affiliation{Graphene Labs, Istituto Italiano di Tecnologia, Genova, Italy}
	
	\author{Camilla Coletti}
	\affiliation{Center for Nanotechnology Innovation@NEST, Istituto Italiano di Tecnologia, Pisa, Italy}
	\affiliation{Graphene Labs, Istituto Italiano di Tecnologia, Genova, Italy}
	
	\author{Tim O. Wehling}
	\affiliation{I. Institute of Theoretical Physics, University of Hamburg, 22607 Hamburg, Germany}
	\affiliation{The Hamburg Centre for Ultrafast Imaging, Luruper Chaussee 149, 22761 Hamburg, Germany}
	
	\author{Isabella Gierz}
	\email[]{isabella.gierz@ur.de}
	\affiliation{Department for Experimental and Applied Physics, University of Regensburg, 93040 Regensburg, Germany}
	
	%Collaboration name if desired (requires use of superscriptaddress
	%option in \documentclass). \noaffiliation is required (may also be
	%used with the \author command).
	%\collaboration can be followed by \email, \homepage, \thanks as well.
	%\collaboration{}
	%\noaffiliation
	
	\date{\today}
	
	\begin{abstract}
	
	Intercalation of epitaxial graphene on SiC(0001) with Sn results in a well-ordered Sn $(1\times1)$ structure on the SiC surface with quasi-freestanding graphene on top. While the electronic properties of the individual layers have been studied in the past, emerging phenomena arising from possible inter-layer interactions between the 2D\,Sn layer and graphene remain unexplored. We use time- and angle-resolved photoemission spectroscopy to reveal a surprisingly short-lived non-equilibrium carrier distribution inside the Dirac cone of Sn-intercalated graphene. Further, we find that the graphene $\pi$-band	exhibits a transient increase in binding energy that we attribute to charging of the graphene layer with holes. We interpret our results with support from density functional theory calculations of the graphene - 2D\,Sn heterostructure that reveal a substantial hybridization between the graphene $\pi$-bands and Sn states, providing a channel for efficient ultrafast charge transfer between the layers. Our results on the graphene - 2D\,Sn model system are expected to trigger similar investigations on related heterostructures obtained by intercalation of epitaxial graphene. Regarding the huge choice of materials that have been successfully intercalated in the past, we believe that the interlayer interactions revealed in the present work only represent the tip of the iceberg with many fascinating emerging phenomena to be discovered in the near future.
	\end{abstract}
	
	% insert suggested keywords - APS authors don't need to do this
	%\keywords{}
	
	%\maketitle must follow title, authors, abstract, and keywords
	\maketitle

% \section{Introduction}
	
Van der Waals (vdW) heterostructures \cite{Geim2013, Novoselov2016} are novel artificial materials with tailored electronic properties that are commonly controlled by the choice of specific materials to be stacked, layer thicknesses and spacings, and the relative orientation of the layers. VdW heterostructures are a possible platform for the experimental realization of various exotic quantum phases \cite{Kennes2021} and may enable new technologies in the fields of flexible electronics, optoelectronics, spintronics, and quantum computing \cite{Liu2016, Jin2018}. They are commonly made by manual cleaving and stacking of layered materials into heterostructures. This procedure works well for proof-of-principle devices. However, sample sizes are small and the interfaces are usually dirty. These problems are resolved by confinement heteroepitaxy \cite{Riedl2009, Gierz2010, Briggs2020} where various kinds of atoms are intercalated at the interface between epitaxial graphene and SiC substrate in ultra-high vacuum. The intercalated atoms self-assemble into well-ordered 2D surface structures on SiC(0001) with quasi-freestanding graphene on top.
	
Here, we focus on a heterostructure consisting of graphene and 2D\,Sn forming a metallic $(1\times1)$ structure on SiC(0001). This system has been previously investigated for the following reasons: Sn intercalation can be used for the fabrication of almost charge-neutral graphene \cite{Kim2016}. Further, the intercalated Sn forms a triangular lattice with exotic electronic properties where Dirac-like and flat bands coexist close to the Fermi level \cite{Hayashi2018, Visikovskiy2018}. Finally, the spin degeneracy of the Sn states at the K-point is lifted due to spin-orbit coupling and broken inversion symmetry resulting in both Zeeman- and Rashba-like spin splittings \cite{Yaji2019, Visikovskiy2018}. Emergent electronic properties that may arise from interlayer interactions between graphene and 2D\,Sn have received little attention so far. Also, non-equilibrium carrier dynamics that is essential for understanding and optimizing the performance of various devices such as transistors, photodetectors, or solar cells remains largely unexplored.

Here, we combine high-resolution angle-resolved photoemission spectroscopy (ARPES) measurements and density functional theory (DFT) calculations to determine the band structure of 2D\,Sn on SiC(0001), confirming the formation of the well-known metallic $(1\times1)$ phase. Next, we use time-resolved ARPES (trARPES) to explore the non-equilibrium carrier dynamics and transient band structure of graphene on 2D\,Sn after photoexcitation with visible pump pulses with a photon energy of $\hbar\omega_{pump}=2$\,eV. We find surprisingly short-lived population dynamics inside the Dirac cone of graphene with lifetimes below $\sim500$\,fs, suggesting carrier relaxation via states of 2D\,Sn. Further, our trARPES results reveal a short-lived transient rigid downshift of the graphene $\pi$-bands that we attribute to a transient charging of the graphene layer with holes. DFT calculations of the graphene - 2D\,Sn heterostructure reveal a strong hybridization between the Dirac cone of graphene and the flat bands consisting of Sn $p_z$ orbitals. The corresponding wave functions are delocalized over both the graphene and the 2D\,Sn layer and mediate ultrafast charge transfer between the layers.

Our results provide the fundamental understanding of ultrafast carrier dynamics in graphene/$(1\times1)$Sn/SiC(0001) relevant for the development of faster and more efficient optoelectronic devices. Also, the graphene - 2D\,Sn interface investigated in the present study is a model system representing a huge class of vdW heterostructures that can be made by confinement heteroepitaxy. Other interfaces between graphene and 2D semiconductors \cite{Gierz2010, Rosenzweig2020, Forti2020}, 2D Mott insulators \cite{Glass2015, Mamiyev2024}, or 2D superconductors \cite{Briggs2020} have been realized experimentally in the past and are expected to reveal intriguing emerging phenomena in the future.

% \section{Methods}

A carbon buffer layer with $(6\sqrt{3}\times6\sqrt{3})R30^{\circ}$ structure was grown by thermal decomposition of hydrogen-etched SiC(0001) in Ar atmosphere \cite{Emtsev2009}. Sn was then deposited from a commercial Knudsen cell at room temperature and intercalated by postannealing at 900\,$^{\circ}$C for 5 minutes. Alternatively, the $(6\sqrt{3}\times6\sqrt{3})R30^{\circ}$ structure was intercalated with hydrogen as described in \cite{Riedl2009} yielding quasi-freestanding monolayer graphene on H-terminated SiC.

High-resolution ARPES experiments were performed at the $1^2$-ARPES end station at Bessy-II \cite{onesquare} with a photon energy of $\hbar\omega=150$\,eV at a sample temperature of $T=13.4$\,K. Energy and angular resolution were $\sim$160\,meV and $\sim$0.5°, respectively.

The trARPES setup was based on a Ti:Sa amplifier (Astrella, Coherent) with a repetition rate of 1\,kHz, a pulse energy of 7\,mJ and a pulse duration of 35\,fs. 5\,mJ of output energy were used to seed an optical parametric amplifier (Topas Twins, Light Conversion) the second harmonic of which was used to generate visible pump pulses with a photon energy of $\hbar\omega_{pump}=2$\,eV and a fluence of $F=0.3$\,mJ/cm$^2$. The remaining 2\,mJ of pulse energy were frequency doubled and used for high harmonics generation in Argon. A single harmonic with a photon energy of $\hbar\omega_{probe}=21.7$\,eV was selected with a grating monochromator. These extreme ultraviolet (XUV) probe pulses were used to eject photoelectrons from the sample. The resulting photoemission current was measured with a hemispherical analyzer (Phoibos 100, Specs) as a function of kinetic energy and emission angle, directly yielding snapshots of the occupied part of the transient band structure of the sample. The energy and temporal resolutions were 190\,meV and 150\,fs, respectively.

We performed the DFT calculations using the Vienna Ab-initio Simulation Package (VASP) \cite{Kresse1993, Kresse1996, Kresse1996b} within the generalized gradient approximation of Perdew, Burke, and Ernzerhof (GGA-PBE) \cite{Perdew1996} and the projector augmented wave (PAW) \cite{Bloechl1994, Kresse1999} basis sets. We obtain the total energy, relaxed structure and electronic structure employing a kinetic energy cutoff of 400\,eV and a $\Gamma$-centered mesh of $12 \times 12 \times 1$ in our calculations. The conjugate gradient algorithm is used for ionic relaxation until all force components are smaller than 0.005\,eV\,\AA$^{-1}$. We do not include spin-orbit coupling in our calculations. In all calculations we keep a fixed lattice constant of $a = 3.095$\,{\AA} for the SiC bulk and Sn$(1\times1)$ phase. For the heterostructure with epitaxial graphene, we adopt a $(\sqrt{3}\times\sqrt{3})R30^{\circ}$ supercell of the Sn/SiC(0001) surface such that graphene is stretched by approx. 15\,\% with a C-C distance of $d_{\mathrm{C}-\mathrm{C}} \approx 1.63$\,\AA. This is a necessary and commonly made approximation \cite{Mattausch2007} to reduce the computational effort. We employed a vacuum of around 6\,{\AA} to eliminate spurious interactions within the periodic cell scheme. The data of our DFT calculations is made available at \cite{url}.

% \section{Results}

\begin{figure}
		\includegraphics[width = 1\columnwidth]{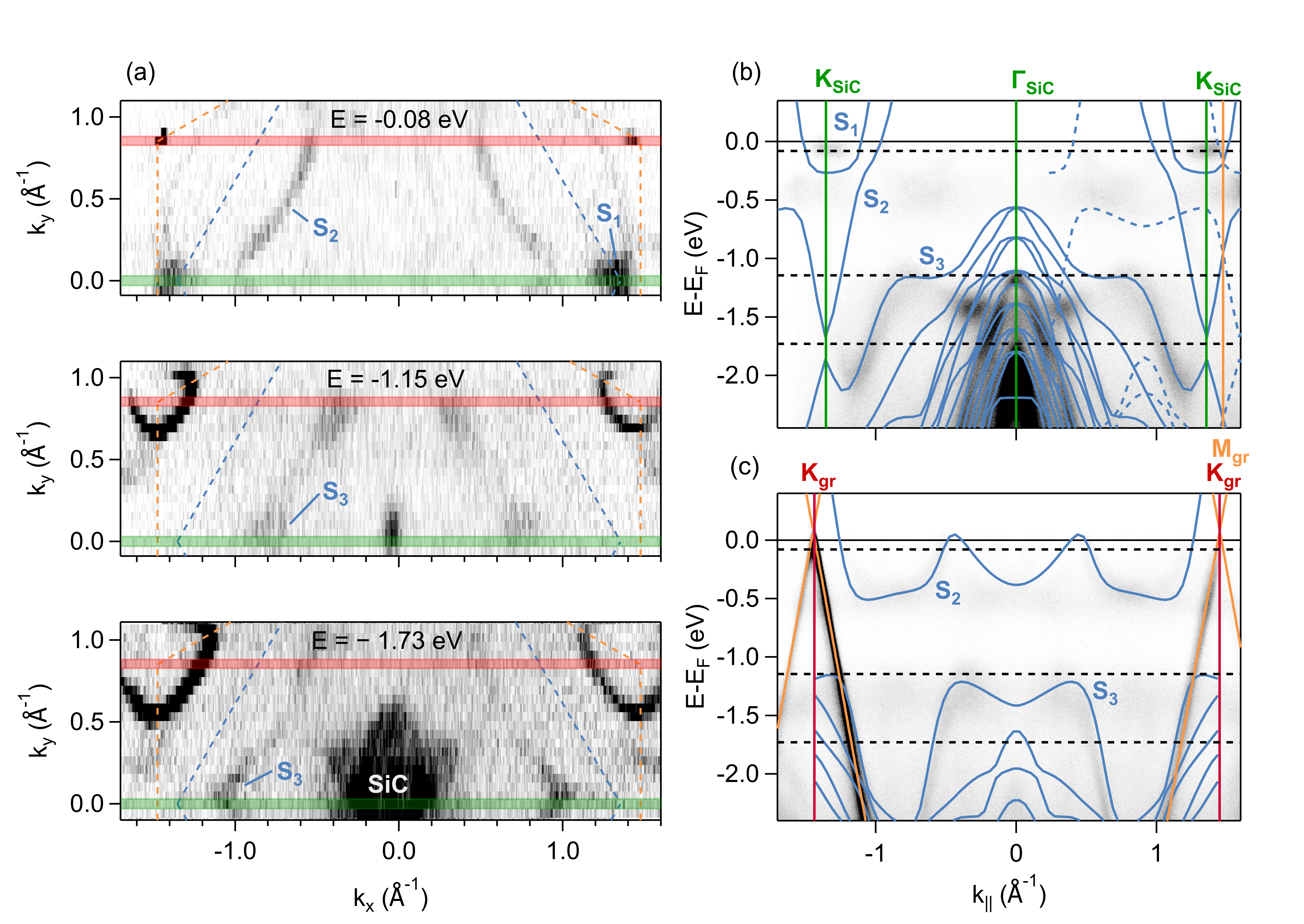}
		\caption{\textbf{High-resolution ARPES of graphene - 2D\,Sn heterostructure.} (a) Photoemission current at constant energy as indicated in each panel. Dashed blue and oranges lines indicated the Brillouin zones of SiC(0001) and graphene, respectively. The three Sn-related states are marked in blue as S$_1$, S$_2$, and S$_3$. Green and red lines indicate the directions along which the dispersion in panels (b) and (c), respectively, are plotted. (b) Band structure measured along the K$\Gamma$K-direction of SiC(0001) together with DFT results of the $(1\times1)$ structure of Sn on SiC(0001) (blue lines). Dashed blue lines are Sn-bands that are backfolded at the M-point of graphene. (c) Band structure measured along the KK-direction of graphene together with DFT results of the $(1\times1)$ structure of Sn on SiC(0001) (blue lines) and free-standing graphene (orange lines). Dashed black lines in (b) and (c) indicate the energies of the constant energy contours in (a).}
		\label{figure1}
	\end{figure}

In Fig. \ref{figure1} we present high-resolution ARPES results for the graphene - 2D\,Sn heterostructure. Figure 1a shows the photoemission current at three selected energies as a function of the two in-plane momenta $k_x$ and $k_y$. The Brillouin zones of graphene and 2D\,Sn are indicated by dashed orange and blue lines, respectively. At $E=-0.08$\,eV, just below the Fermi level, we observe small dots at the corners of the graphene Brillouin zone originating from the almost charge neutral Dirac cone of graphene. We further find two Sn-related electron pockets labeled S$_1$ and S$_2$, respectively. S$_1$ is centered around the K-points of the hexagonal Brillouin zone of 2D\,Sn. S$_2$ exhibits the shape of a distorted hexagon centered around the $\Gamma$-point with corners pointing towards the K-points of 2D\,Sn. At lower energies ($E=-1.15$\,eV and $E=-1.73$\,eV) the Dirac cone of graphene exhibits its typical triangular shape. The hexagon with the corners pointing towards the K-points of the Brillouin zone of 2D\,Sn belongs to another Sn-related state labeled S$_3$. The intensity around the $\Gamma$-point is attributed to states belonging to the SiC substrate.

The dispersion of the electronic states along the green and red lines in Fig. 1a is shown in Figs. 1b and c, respectively, together with DFT calculations for a Sn $(1\times1)$ structure on SiC(0001) (blue lines) and free-standing graphene (orange lines). The calculations were shifted by $-290$\,meV (2D\,Sn) and $+80$\,meV (graphene) to account for the experimentally observed doping level. In addition, the DFT calculations for free-standing graphene were stretched by a factor of 1.18 in energy to account for the experimentally observed Fermi velocity $v_F$. The overall agreement between theory and experiment, as well as with existing literature \cite{Kim2016, Hayashi2018, Visikovskiy2018, Yaji2019} is excellent and provides direct evidence that the intercalated Sn forms the well-known metallic $(1\times1)$ structure on SiC(0001).
 
\begin{figure}
		\includegraphics[width = 1\columnwidth]{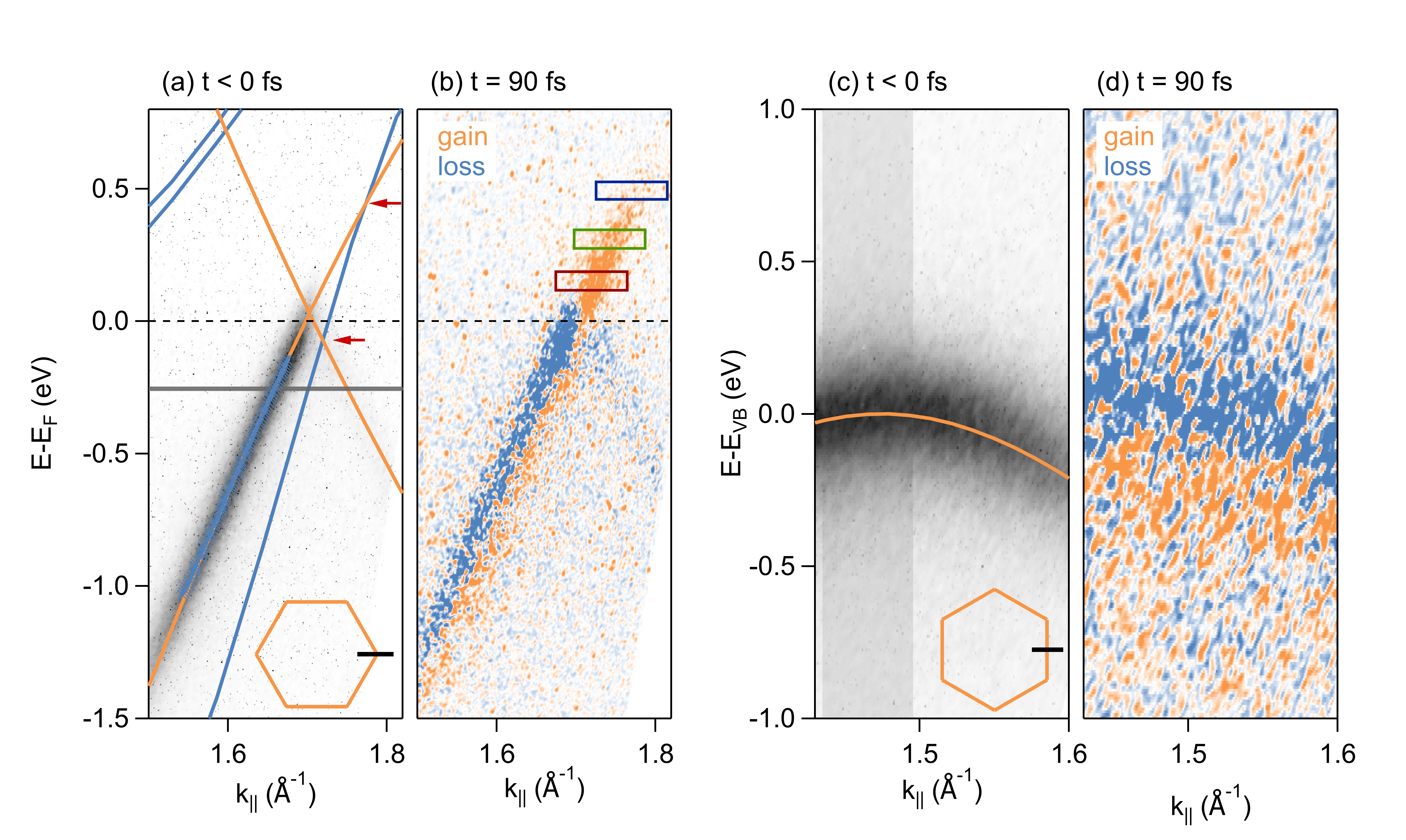}
		\caption{\textbf{Time-resolved ARPES of graphene - 2D\,Sn heterostructure.} (a) Photoemission current along the $\Gamma$K-direction of graphene at the K-point of graphene (see Brillouin zone sketch in the inset) measured before the arrival of the pump pulse together with DFT calculations for freestanding graphene (orange lines) and $(1\times1)$Sn/SiC(0001) (blue lines). Red arrows mark the intersections of the individual DFT band structures. The thick gray line indicates the energy where the MDCs in Fig. \ref{figure3}a were extracted. The blue data points are the result of MDC fits in Fig. \ref{figure3}a. (b) Pump-induced changes of the photoemission current 90\,fs after photo-excitation with a visible pump pulse with a photon energy of $\hbar\omega_{pump}=2$\,eV and a fluence of $F=0.3$\,mJ/cm$^2$. Orange and blue indicate a gain and loss, respectively, of photoelectrons with respect to negative pump-probe delay. Colored boxes indicate the areas of integration for the pump-probe traces presented in Fig. \ref{figure4}a. (c) and (d) are the same as (a) and (b) but measured at the M-point of graphene (see inset in c). The gray-shaded area in (c) indicates the momentum range where the EDCs in Fig. \ref{figure3}c were extracted.}
	\label{figure2}
	\end{figure}

After having confirmed the structure of our sample, we now investigate the non-equilibrium carrier dynamics triggered by photoexcitation with a visible pump pulse with a photon energy of $\hbar\omega_{pump}=2$\,eV and a fluence of $F=0.3$\,mJ/cm$^2$. Figure \ref{figure2} shows trARPES snapshots of the graphene $\pi$-band at the K (Fig. \ref{figure2}a and b) and M-point (Fig. \ref{figure2}c and d) of the graphene Brillouin zone. The Sn-related bands could not be detected in the trARPES signal possibly due to a poor cross section at the photon energy used and/or the limited signal-to-noise ratio. 

Figure \ref{figure2}a shows a snapshot of the Dirac cone taken at negative pump-probe delay before the arrival of the pump pulse together with DFT calculations for free-standing graphene (orange lines) and the Sn $(1\times1)$ structure on SiC(0001) (blue lines). The corresponding pump-induced changes of the photoemission current at a pump-probe delay of $t=90$\,fs are shown in Fig. \ref{figure2}b. Gain and loss of photoelectrons with respect to negative delay are shown in orange and blue, respectively. We find a pronounced gain signal above the Fermi level, indicating a transient population of the conduction band up to energies of $\sim0.5$\,eV above the Fermi level. The valence band shows a loss that is strongest close to the Fermi level but reaches to much lower energies beyond the energy window covered in the experiment. The loss signal is accompanied by a pronounced gain below the equilibrium position of the Dirac cone, suggesting a transient down-shift of the dispersion. This is corroborated by the trARPES signal of the $\pi$-band at M in Figs. \ref{figure2}c and d, that show a snapshot at negative delay and the pump-induced changes of the photoemission current at a pump-probe delay of $t=90$\,fs, respectively. Just like Fig. \ref{figure2}b, Fig. \ref{figure2}d also shows a pronounced loss of photoemission current at the equilibrium position of the $\pi$-band and a gain below, consistent with a rigid down-shift of the $\pi$-band.

\begin{figure}
		\includegraphics[width = 0.7\columnwidth]{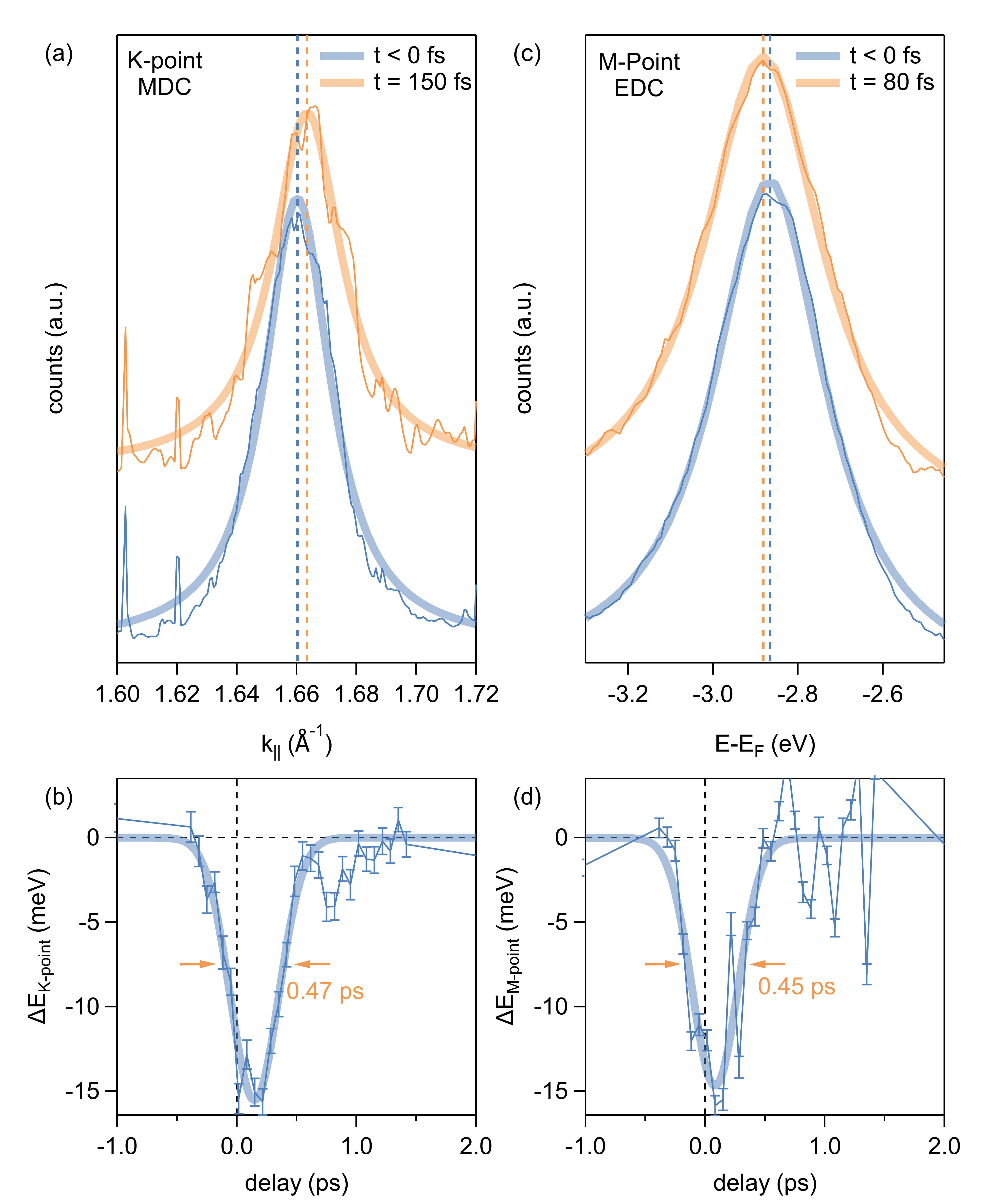}
		\caption{\textbf{Transient shift of graphene $\pi$-band.} (a) MDCs extracted along the thick gray line in Fig. \ref{figure1}a for two different time delays together with Lorentzian fits. (b) Time-dependent peak shift of the Dirac cone from (a) together with Gaussian fit with a full width at half maximum of $470\pm30$\,fs. The MDCs in (a) are offset for clarity. (c) Energy distribution curves obtained by integrating the ARPES snapshots at the M-point over the momentum range marked by the gray-shaded area in Fig. \ref{figure2}c together with Lorentzian fits for two different pump-probe delays. (d) Time-dependent peak position of the Lorentzian fits from (c) together with Gaussian fit with a full width at half maximum of $450\pm70$\,fs.}
	\label{figure3}
	\end{figure}

To analyze the pump-probe signal in greater detail, we first explore the possible down-shift of the $\pi$-band suggested by the trARPES snapshots in Fig. \ref{figure2}. To determine the transient binding energy of the Dirac cone at the K-point of the graphene Brillouin zone we extract momentum distribution curves (MDCs) at constant energy from the trARPES snapshot in Fig. \ref{figure2}a, that we fit with a Lorentzian (an example is shown in Fig. \ref{figure3}a). The resulting peak position $k$ as a function of energy $E$ yields the dispersion relation of the Dirac cone $E(k)$ that we fit with a straight line. We repeat this procedure for all pump-probe delays, fix the slope of the linear fit to the average value for negative pump-probe delays ($\hbar v_F=6.8$\,eV\AA), and obtain the transient shift of the Dirac cone that is shown in Fig. \ref{figure3}b. The blue line in Fig. \ref{figure3}b represents a Gaussian fit that reproduces the fact that the peak shift seems to be symmetric about its  minimum. We find a transient downshift of the Dirac cone of $\Delta E_K=-16\pm1$\,meV with a Gaussian full width at half maximum of FWHM$=470\pm30$\,fs. The transient binding energy of the $\pi$-band at the M-point is extracted as follows: We integrate the trARPES snapshot in Fig. \ref{figure2}c over the momentum range indicated by the gray-shaded area and fit the resulting energy distribution curve (EDC) with a Lorentzian peak and a Shirley background (see Fig. \ref{figure3}c). We then repeat this procedure for all other pump-probe delays. The temporal evolution of the change of the Lorentzian peak position is shown in Fig. \ref{figure3}d together with a Gaussian fit (blue line). We find a transient downshift of the $\pi$-band at M of $\Delta E_M=-15\pm2$\,meV with a Gaussian full width at half maximum of FWHM$=450\pm70$\,fs. The shifts of the $\pi$-band at K and M are the same within the error bars in terms of amplitude and lifetime, indicating a rigid down-shift of the whole $\pi$-band.

\begin{figure}
		\includegraphics[width = 1\columnwidth]{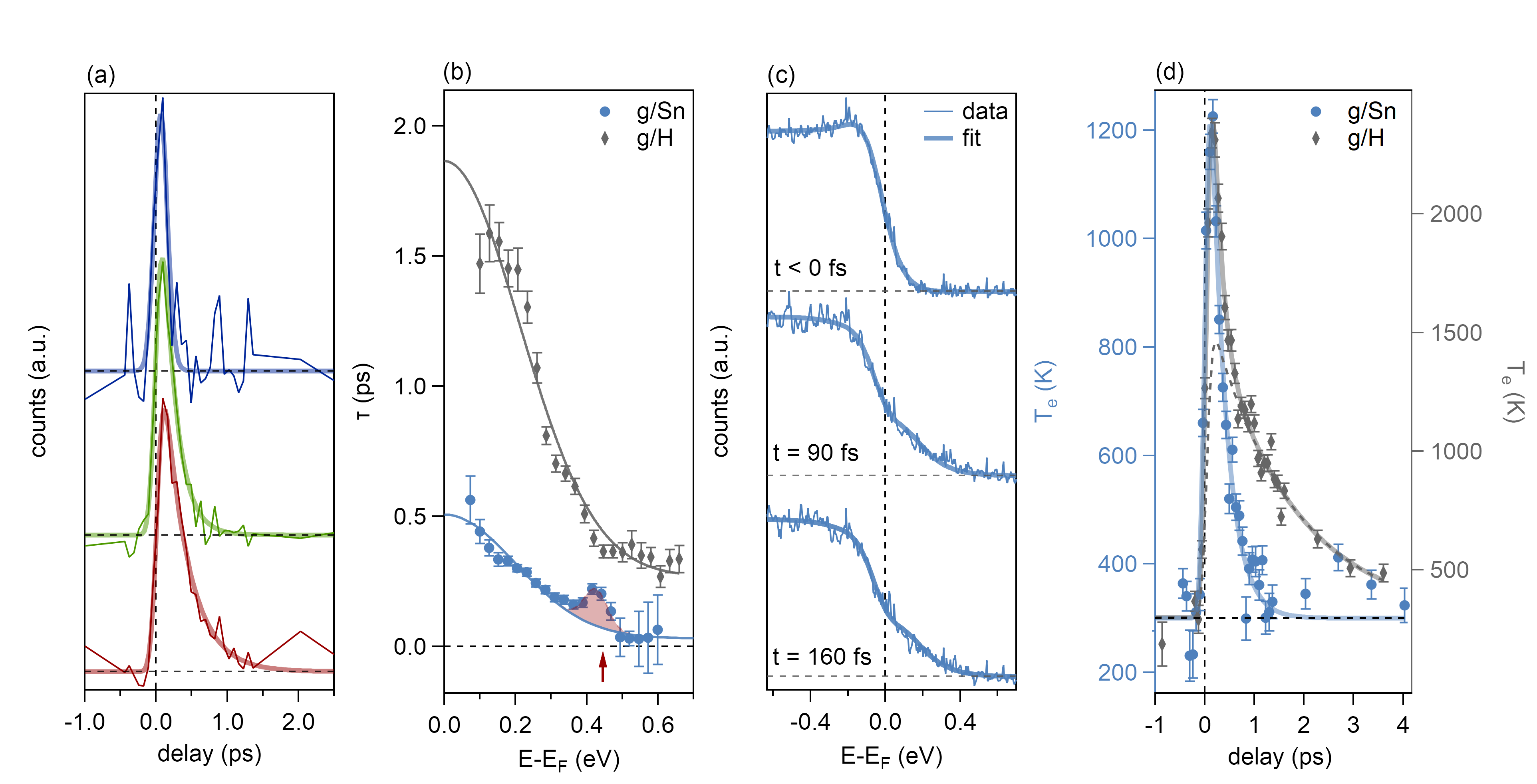}
		\caption{\textbf{Non-equilibrium population dynamics of graphene's Dirac cone} (a) Photoemission current intergrated over the areas marked by the colored boxes in Fig. \ref{figure2}b as a function of pump-probe delay together with single-exponential fits. (b) Exponential lifetime $\tau$ extracted from the fits in (a) as a function of energy. Continuous lines serve as guide to the eye. (c) Distribution of electrons inside the Dirac cone of graphene obtained by integrating ARPES snapshots over the momentum range indicated by the gray-shaded area in Fig. \ref{figure2}a for three different pump-probe delays as indicated. Thick lines are Fermi-Dirac fits as described in the main text. (d) Transient electronic temperature of the carriers inside the Dirac cone together with exponential fits. Blue and gray data points in (b) and (d) belong to graphene on 2D\,Sn and quasi-free-standing graphene on H-SiC, respectively. The dashed gray line in (d) highlights the second slow decay component that is only present in the case of quasi-freestanding graphene. }
	\label{figure4}
	\end{figure}

Next, we focus on the energy- and momentum-resolved non-equilibrium carrier dynamics. Figure \ref{figure4}a shows the photoemission current, integrated over the areas marked by the colored boxes in Fig. \ref{figure2}b, as a function of pump-probe delay together with single-exponential fits. The resulting lifetime $\tau$ as a function of energy is shown in blue in Fig. \ref{figure4}b. We restrict ourselves to energies $E-E_F>0$, as the signal-to-noise ratio for $E-E_F<0$ is significantly lower. We find that the lifetime increases from $\tau\sim60$\,fs at $E-E_F=0.6$\,eV to $\tau\sim500$\,fs close to the Fermi level. In addition, the lifetime is found to exhibit a small peak around $E-E_F=0.42$\,eV (red-shaded area in Fig. \ref{figure4}b) where the graphene and 2D\,Sn DFT band structures in Fig. \ref{figure2}a are predicted to cross. Figure \ref{figure4}c shows the transient carrier distribution inside the Dirac cone as a function of energy. The data points were obtained by integrating the trARPES snapshot over the momentum range in Fig. \ref{figure2}a. These carrier distributions were fitted with a function that takes into account the linear density of states of the Dirac cone, the Fermi Dirac distribution, the finite energy resolution as well as the finite momentum resolution that arises from the finite width of the entrance slit of the hemispherical analyzer (blue lines in Fig. \ref{figure4}c). The resulting electronic temperature is shown in Fig. \ref{figure4}d together with a single-exponential fit. The electronic temperature is found to reach a peak value of $\sim1200$\,K and to cool down with a lifetime of $\tau=300\pm30$\,fs.

For comparison, Figs. \ref{figure4}b and d show data for quasi-freestanding graphene on H-SiC (gray data points). We find that both the population lifetimes in Fig. \ref{figure4}b and the peak electronic temperature in Fig. \ref{figure4}d are larger in quasi-freestanding graphene compared to graphene on 2D\,Sn by a factor of $\sim3$ and $\sim2$, respectively. Further, in quasi-freestanding graphene, the peak in the population lifetime around 0.42\,eV is found to be absent and the electronic temperature is found to cool with a double-exponential rather than a single-exponential decay with lifetimes of $\tau_1=170\pm50$\,fs and $\tau_2=1.7\pm0.2$\,ps.

%\section{Discussion}

The non-equilibrium carrier distribution established in photo-doped graphene is known to relax as follows: Within $\sim30$\,fs a hot Fermi-Dirac distribution is established \cite{Breusing2009} that subsequently cools down by the emission of optical and acoustic phonons with characteristic timescales on the order of a few hundreds of femtoseconds and some picoseconds, respectively \cite{Kampfrath2005, Yan2009, Johannsen2013}. The gradual cooling of the hot Fermi-Dirac distribution directly results in energy-resolved population lifetimes $\tau(E)$ that increase when approaching the Fermi level \cite{GierzFD2014, GierzJPCM2015, Johannsen2015}. Our results for free-standing graphene for both $\tau(E)$ and $T_e(t)$ nicely agree with previous results on similar samples from literature \cite{Johannsen2013, GierzFD2014, Ulstrup2015, Johannsen2015}.

Our results for the non-equilibrium carrier dynamics of graphene on 2D\,Sn, however, deviate from this picture in several ways: (1) The peak electronic temperature for graphene on 2D\,Sn is a factor of $\sim2$ smaller than for quasi-freestanding graphene for the same excitation conditions. Both samples are lightly p-doped \cite{Riedl2009}, suggesting that the absorption coefficient for the incident pump photons should be similar. Therefore, we attribute the difference in peak electronic temperature to the presence of the metallic 2D\,Sn layer that screens the electric field of the pump pulses. (2) In graphene on 2D\,Sn, the population lifetimes $\tau(E)$ are a factor of $\sim3$ smaller and the electronic temperature cools faster compared to quasi-freestanding graphene. This suggests that the photoexcited carriers inside the Dirac cone of graphene on 2D\,Sn efficiently relax via states provided by the metallic 2D\,Sn layer. A similar acceleration of carrier relaxation due to the presence of a metallic substrate has been previously observed for other sample systems in Refs. \cite{Ulstrup2015, Majchrzak2021}. (3) The population lifetimes for graphene on 2D\,Sn exhibit an additional peak around $E-E_F=0.42$\,eV that is absent in quasi-freestanding graphene. This peak is very close in energy to the crossing point of the DFT band structures of free-standing graphene and Sn$(1\times1)$/SiC in Fig. \ref{figure2}b, suggesting a possible influence of the 2D\,Sn layer on the pump-probe signal. (4) Graphene on 2D\,Sn shows a transient increase in binding energy. This suggests that the graphene layer becomes positively charged. The most likely origin of the extra holes in graphene is the 2D\,Sn layer. 

\begin{figure}
		\includegraphics[width = 0.7\columnwidth]{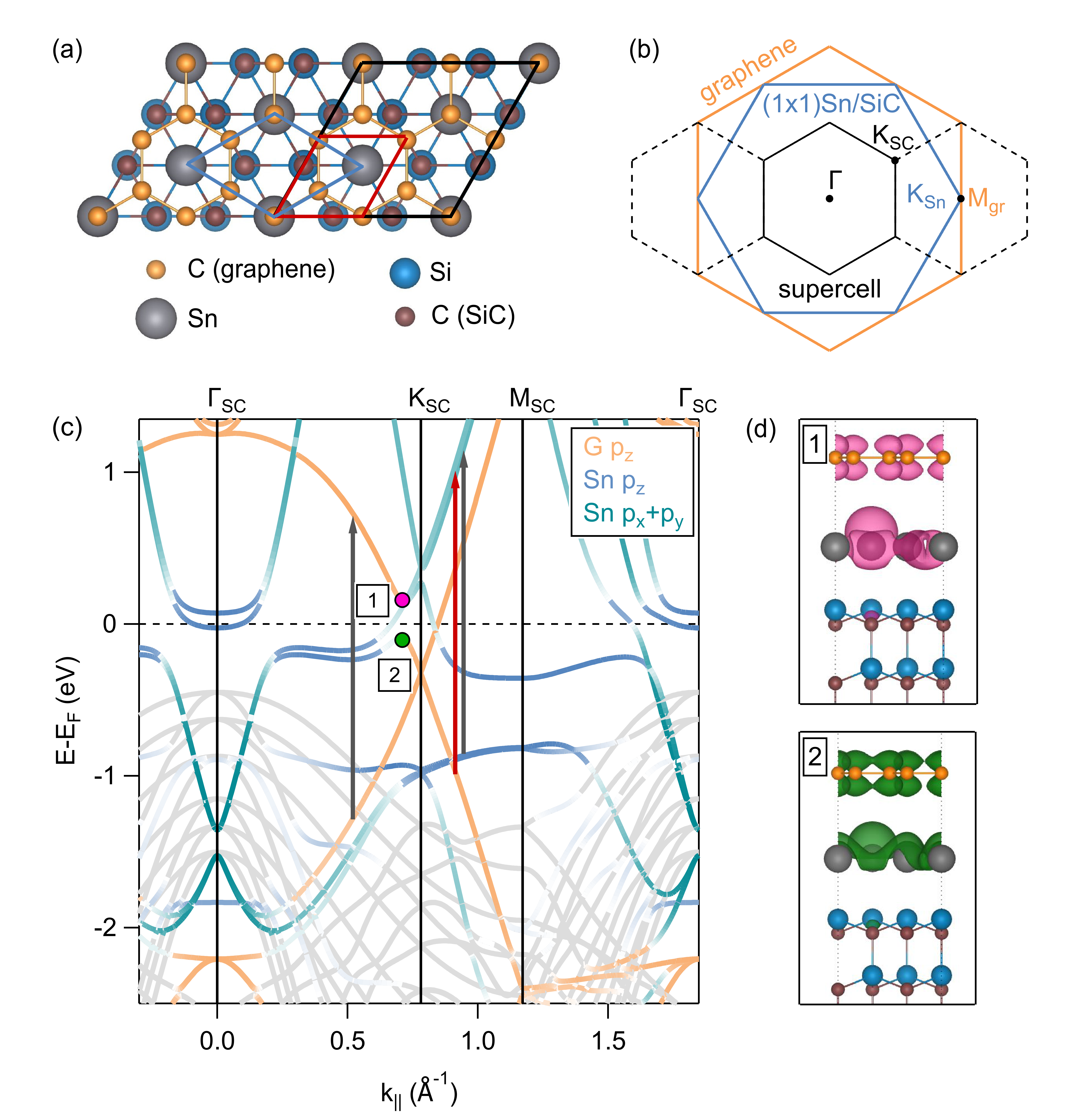}
		\caption{\textbf{DFT calculations for graphene - 2D\,Sn heterostructure.} (a) Structural model with $(\sqrt{3}\times\sqrt{3})R30^{\circ}$ unit cell. (b) Sketch of the Brillouin zones of graphene (orange), $(1\times1)$Sn/SiC(0001) (blue) and $(\sqrt{3}\times\sqrt{3})R30^{\circ}$ supercell (black). (c) Calculated band structure along $\Gamma$KM$\Gamma$ of the Brillouin zone of the $(\sqrt{3}\times\sqrt{3})R30^{\circ}$ supercell (SC). The color-code indicates the orbital composition of the states. Red and gray arrows indicate possible inter-layer and intra-layer transitions triggered by photo-excitation at $\hbar\omega_{pump}=2$\,eV. The pink and green dots labeled 1 and 2 correspond to the location in the band structure $E(k)$ where the two electron densities shown in panel (d) were computed. (d) Absolute square of the wave function corresponding to the two eigenstates marked by pink and green dots in (c).}
	\label{figure5}
	\end{figure}

In order to provide a microscopic explanation for the observed interlayer interactions, we compute the band structure of the graphene - 2D\,Sn heterostructure with DFT. The employed $(\sqrt{3}\times\sqrt{3})R30^{\circ}$ supercell and corresponding Brillouin zone are sketched in Figs. \ref{figure5}a and b, respectively. In Fig. \ref{figure5}c we display the computed band structure. The color code reveals the orbital composition of the states. Sn S$_1$, S$_2$ and S$_3$ states originating from Sn $(p_x+p_y)$-, and $p_z$-orbitals are shown in green and blue, respectively. The graphene $\pi$-band is shown in orange. Despite the weak vdW coupling between graphene and 2D\,Sn we find pronounced deviations from a mere sum of the individual band structures. There is a strong hybridization between graphene and Sn states at the Fermi level along the $\Gamma$K-direction close to the K-point of the Brillouin zone of the supercell. We find a pronounced band gap opening of $E_{gap}\sim230$\,meV in the Dirac cone and a simultaneous change of the orbital composition from Sn-like to graphene-like. 

For a direct comparison between theory and experimental data, the K-point of SiC (and the M-point of graphene) needs to be mapped onto the $\Gamma$-point of the Brillouin zone of the $(\sqrt{3}\times\sqrt{3})R30^{\circ}$ supercell (see Fig. \ref{figure5}b). We find that the dispersion of the Sn states agrees well with our high-resolution ARPES data in Fig. \ref{figure1} including the same $-290$\,meV shift as in Fig. \ref{figure1}. The Dirac cone of graphene, however, exhibits a pronounced n-doping in the DFT band structure in Fig. \ref{figure5}c, whereas it is slightly p-doped in the experimental data in Figs. \ref{figure1} and \ref{figure2}a,b. This discrepancy also affects the intersection (and possible hybridization) points of the graphene and 2D\,Sn dispersions that is found to occur at the Fermi level in Fig. \ref{figure5}c but likely occurs around $0.44$\,eV above the Fermi level in the experimental data in Fig. \ref{figure2}a. While the existence of the hybridization gap in the DFT band structure of the supercell is found to be robust against sensible structural changes including interlayer distance, strain, and local stacking order, its size is found to be strongly dependent on the interlayer distance. Therefore, the size of the actual band gap might be smaller than the value predicted by the DFT calculations which explains why the gap is not observable in the trARPES data in Fig. \ref{figure2}b with an energy resolution of 190\,meV. The fact that the population lifetimes in Fig. \ref{figure4}b show a peak close to the energy where the graphene and Sn states are predicted to intersect, might be an indication for a band gap opening in the Dirac cone.

In Fig. \ref{figure5}d we plot the electron density for the states marked by the pink and green dots in Fig. \ref{figure5}c. We find that, at the momentum where the band gap in the Dirac cone opens, the electron density is delocalized over both the graphene and the 2D\,Sn layer. This provides a natural channel for efficient ultrafast charge transfer between the two layers \cite{Hofmann2023}.

Based on these findings we propose two possible scenarios for the transient charging of the graphene layer. The first scenario involves a direct electronic transition exciting electrons from occupied states in the graphene Dirac cone into unoccupied Sn $p_x$- and $p_y$-states (red arrow in Fig. \ref{figure5}c). This excitation would leave the graphene layer positively charged and the 2D\,Sn layer negatively charged. In the second scenario, electronic excitations would mainly occur within the individual layers (gray arrows in Fig. \ref{figure5}c). Ultrafast electron transfer from graphene to 2D\,Sn could then occur via the charge transfer states in Fig. \ref{figure5}d. Further investigations are needed to decide which of the two scenarios is more likely or if a combination of the two scenarios applies.

% \section{Summary}

In summary, our results clearly indicate that --- despite the weak vdW interaction between the individual layers --- graphene - 2D\,Sn heterostructures exhibit new emerging electronic properties that cannot be described by the mere sum of the individual layers. Using ARPES and DFT we confirmed that Sn intercalated at the interface between epitaxial graphene and SiC(0001) substrate arranges into the well known metallic $(1\times1)$ structure. Using trARPES we have shown that the non-equilibrium carrier distribution established in the Dirac cone of graphene via photo-doping with visible pump pulses rapidly decays via states of the 2D\,Sn layer. We have further demonstrated that the graphene layer exhibits a transient charging with holes. Supported by DFT calculations that reveal a pronounced hybridization between the graphene Dirac cone and Sn $p_z$-states close to the Fermi level, we attribute this either to direct interlayer photoexcitation, or intralayer photoexcitation followed by ultrafast charge separation via charge transfer states, or a combination of both. Future calculations of the absorption and charge transfer rates for the graphene/$(1\times1)$Sn/SiC(0001) structure should clarify which of these scenarios is the most likely one.

Our results have important implications for the design of future ultrafast optoelectronic devices such as sensors or photodetectors. Further, our results will pave the way towards similar investigations on other graphene-based heterostructures made by confinement heteroepitaxy with more intriguing emergent phenomena likely to be found in the near future.

\section{Acknowledgments}
We thank Dr. Emile Rienks for support during the ARPES measurements at the BESSY II electron storage ring operated by the Helmholtz-Zentrum Berlin. This work received funding from the Deutsche Forschungsgemeinschaft (DFG) via the research unit FOR 5242 (project No. 449119662), the collaborative research center CRC 1277 (project No. 314695032), the Cluster of Excellence ‘CUI: Advanced Imaging of Matter’ - EXC 2056 - project ID 390715994, as well as from the European Union’s Horizon 2020 research and innovation program under Grant Agreement No. 851280-ERC-2019-STG and Graphene Core 3 project (Grant Agreement No. 881603).

\clearpage

\pagebreak

\bibliography{literature_IGP}% Produces the bibliography via BibTeX.

%apsrev4-2.bst 2019-01-14 (MD) hand-edited version of apsrev4-1.bst
%Control: key (0)
%Control: author (8) initials jnrlst
%Control: editor formatted (1) identically to author
%Control: production of article title (0) allowed
%Control: page (0) single
%Control: year (1) truncated
%Control: production of eprint (0) enabled
\begin{thebibliography}{36}%
\makeatletter
\providecommand \@ifxundefined [1]{%
 \@ifx{#1\undefined}
}%
\providecommand \@ifnum [1]{%
 \ifnum #1\expandafter \@firstoftwo
 \else \expandafter \@secondoftwo
 \fi
}%
\providecommand \@ifx [1]{%
 \ifx #1\expandafter \@firstoftwo
 \else \expandafter \@secondoftwo
 \fi
}%
\providecommand \natexlab [1]{#1}%
\providecommand \enquote  [1]{``#1''}%
\providecommand \bibnamefont  [1]{#1}%
\providecommand \bibfnamefont [1]{#1}%
\providecommand \citenamefont [1]{#1}%
\providecommand \href@noop [0]{\@secondoftwo}%
\providecommand \href [0]{\begingroup \@sanitize@url \@href}%
\providecommand \@href[1]{\@@startlink{#1}\@@href}%
\providecommand \@@href[1]{\endgroup#1\@@endlink}%
\providecommand \@sanitize@url [0]{\catcode `\\12\catcode `\$12\catcode
  `\&12\catcode `\#12\catcode `\^12\catcode `\_12\catcode `\%12\relax}%
\providecommand \@@startlink[1]{}%
\providecommand \@@endlink[0]{}%
\providecommand \url  [0]{\begingroup\@sanitize@url \@url }%
\providecommand \@url [1]{\endgroup\@href {#1}{\urlprefix }}%
\providecommand \urlprefix  [0]{URL }%
\providecommand \Eprint [0]{\href }%
\providecommand \doibase [0]{https://doi.org/}%
\providecommand \selectlanguage [0]{\@gobble}%
\providecommand \bibinfo  [0]{\@secondoftwo}%
\providecommand \bibfield  [0]{\@secondoftwo}%
\providecommand \translation [1]{[#1]}%
\providecommand \BibitemOpen [0]{}%
\providecommand \bibitemStop [0]{}%
\providecommand \bibitemNoStop [0]{.\EOS\space}%
\providecommand \EOS [0]{\spacefactor3000\relax}%
\providecommand \BibitemShut  [1]{\csname bibitem#1\endcsname}%
\let\auto@bib@innerbib\@empty
%</preamble>
\bibitem [{\citenamefont {Geim}\ and\ \citenamefont
  {Grigorieva}(2013)}]{Geim2013}%
  \BibitemOpen
  \bibfield  {author} {\bibinfo {author} {\bibfnamefont {A.~K.}\ \bibnamefont
  {Geim}}\ and\ \bibinfo {author} {\bibfnamefont {I.~V.}\ \bibnamefont
  {Grigorieva}},\ }\bibfield  {title} {\bibinfo {title} {{Van der Waals
  heterostructures}},\ }\href@noop {} {\bibfield  {journal} {\bibinfo
  {journal} {Nature}\ }\textbf {\bibinfo {volume} {499}},\ \bibinfo {pages}
  {419} (\bibinfo {year} {2013})},\ \Eprint
  {https://arxiv.org/abs/https://www.nature.com/articles/nature12385}
  {https://www.nature.com/articles/nature12385} \BibitemShut {NoStop}%
\bibitem [{\citenamefont {Novoselov}\ \emph {et~al.}(2016)\citenamefont
  {Novoselov}, \citenamefont {Mishchenko}, \citenamefont {Carvalho},\ and\
  \citenamefont {Neto}}]{Novoselov2016}%
  \BibitemOpen
  \bibfield  {author} {\bibinfo {author} {\bibfnamefont {K.~S.}\ \bibnamefont
  {Novoselov}}, \bibinfo {author} {\bibfnamefont {A.}~\bibnamefont
  {Mishchenko}}, \bibinfo {author} {\bibfnamefont {A.}~\bibnamefont
  {Carvalho}},\ and\ \bibinfo {author} {\bibfnamefont {A.~H.~C.}\ \bibnamefont
  {Neto}},\ }\bibfield  {title} {\bibinfo {title} {{2D materials and van der
  Waals heterostructures}},\ }\href {https://doi.org/10.1126/science.aac9439}
  {\bibfield  {journal} {\bibinfo  {journal} {Science}\ }\textbf {\bibinfo
  {volume} {353}},\ \bibinfo {pages} {aac9439} (\bibinfo {year} {2016})},\
  \Eprint
  {https://arxiv.org/abs/https://www.science.org/doi/pdf/10.1126/science.aac9439}
  {https://www.science.org/doi/pdf/10.1126/science.aac9439} \BibitemShut
  {NoStop}%
\bibitem [{\citenamefont {Kennes}\ \emph {et~al.}(2021)\citenamefont {Kennes},
  \citenamefont {Claassen}, \citenamefont {Xian}, \citenamefont {Georges},
  \citenamefont {Millis}, \citenamefont {Hone}, \citenamefont {Dean},
  \citenamefont {Basov}, \citenamefont {Pasupathy},\ and\ \citenamefont
  {Rubio}}]{Kennes2021}%
  \BibitemOpen
  \bibfield  {author} {\bibinfo {author} {\bibfnamefont {D.~M.}\ \bibnamefont
  {Kennes}}, \bibinfo {author} {\bibfnamefont {M.}~\bibnamefont {Claassen}},
  \bibinfo {author} {\bibfnamefont {L.}~\bibnamefont {Xian}}, \bibinfo {author}
  {\bibfnamefont {A.}~\bibnamefont {Georges}}, \bibinfo {author} {\bibfnamefont
  {A.~J.}\ \bibnamefont {Millis}}, \bibinfo {author} {\bibfnamefont
  {J.}~\bibnamefont {Hone}}, \bibinfo {author} {\bibfnamefont {C.~R.}\
  \bibnamefont {Dean}}, \bibinfo {author} {\bibfnamefont {D.~N.}\ \bibnamefont
  {Basov}}, \bibinfo {author} {\bibfnamefont {A.~N.}\ \bibnamefont
  {Pasupathy}},\ and\ \bibinfo {author} {\bibfnamefont {A.}~\bibnamefont
  {Rubio}},\ }\bibfield  {title} {\bibinfo {title} {{Moir\'e heterostructures
  as a condensed-matter quantum simulator}},\ }\href@noop {} {\bibfield
  {journal} {\bibinfo  {journal} {Nature Physics}\ }\textbf {\bibinfo {volume}
  {17}},\ \bibinfo {pages} {155} (\bibinfo {year} {2021})},\ \Eprint
  {https://arxiv.org/abs/https://www.nature.com/articles/s41567-020-01154-3}
  {https://www.nature.com/articles/s41567-020-01154-3} \BibitemShut {NoStop}%
\bibitem [{\citenamefont {Liu}\ \emph {et~al.}(2016)\citenamefont {Liu},
  \citenamefont {Weiss}, \citenamefont {Duan}, \citenamefont {Cheng},
  \citenamefont {Huang},\ and\ \citenamefont {Duan}}]{Liu2016}%
  \BibitemOpen
  \bibfield  {author} {\bibinfo {author} {\bibfnamefont {Y.}~\bibnamefont
  {Liu}}, \bibinfo {author} {\bibfnamefont {N.~O.}\ \bibnamefont {Weiss}},
  \bibinfo {author} {\bibfnamefont {X.}~\bibnamefont {Duan}}, \bibinfo {author}
  {\bibfnamefont {H.-C.}\ \bibnamefont {Cheng}}, \bibinfo {author}
  {\bibfnamefont {Y.}~\bibnamefont {Huang}},\ and\ \bibinfo {author}
  {\bibfnamefont {X.}~\bibnamefont {Duan}},\ }\bibfield  {title} {\bibinfo
  {title} {{Van der Waals heterostructures and devices}},\ }\href@noop {}
  {\bibfield  {journal} {\bibinfo  {journal} {Nature Reviews Materials}\
  }\textbf {\bibinfo {volume} {1}},\ \bibinfo {pages} {16042} (\bibinfo {year}
  {2016})},\ \Eprint
  {https://arxiv.org/abs/https://www.nature.com/articles/natrevmats201642}
  {https://www.nature.com/articles/natrevmats201642} \BibitemShut {NoStop}%
\bibitem [{\citenamefont {Jin}\ \emph {et~al.}(2018)\citenamefont {Jin},
  \citenamefont {Ma}, \citenamefont {Karni}, \citenamefont {Regan},
  \citenamefont {Wang},\ and\ \citenamefont {Heinz}}]{Jin2018}%
  \BibitemOpen
  \bibfield  {author} {\bibinfo {author} {\bibfnamefont {C.}~\bibnamefont
  {Jin}}, \bibinfo {author} {\bibfnamefont {E.~Y.}\ \bibnamefont {Ma}},
  \bibinfo {author} {\bibfnamefont {O.}~\bibnamefont {Karni}}, \bibinfo
  {author} {\bibfnamefont {E.~C.}\ \bibnamefont {Regan}}, \bibinfo {author}
  {\bibfnamefont {F.}~\bibnamefont {Wang}},\ and\ \bibinfo {author}
  {\bibfnamefont {T.~F.}\ \bibnamefont {Heinz}},\ }\bibfield  {title} {\bibinfo
  {title} {{Ultrafast dynamics in van der Waals heterostructures}},\
  }\href@noop {} {\bibfield  {journal} {\bibinfo  {journal} {Nature
  Nanotechnology}\ }\textbf {\bibinfo {volume} {13}},\ \bibinfo {pages} {994}
  (\bibinfo {year} {2018})},\ \Eprint
  {https://arxiv.org/abs/https://www.nature.com/articles/s41565-018-0298-5}
  {https://www.nature.com/articles/s41565-018-0298-5} \BibitemShut {NoStop}%
\bibitem [{\citenamefont {Riedl}\ \emph {et~al.}(2009)\citenamefont {Riedl},
  \citenamefont {Coletti}, \citenamefont {Iwasaki}, \citenamefont {Zakharov},\
  and\ \citenamefont {Starke}}]{Riedl2009}%
  \BibitemOpen
  \bibfield  {author} {\bibinfo {author} {\bibfnamefont {C.}~\bibnamefont
  {Riedl}}, \bibinfo {author} {\bibfnamefont {C.}~\bibnamefont {Coletti}},
  \bibinfo {author} {\bibfnamefont {T.}~\bibnamefont {Iwasaki}}, \bibinfo
  {author} {\bibfnamefont {A.~A.}\ \bibnamefont {Zakharov}},\ and\ \bibinfo
  {author} {\bibfnamefont {U.}~\bibnamefont {Starke}},\ }\bibfield  {title}
  {\bibinfo {title} {{Quasi-Free-Standing Epitaxial Graphene on SiC Obtained by
  Hydrogen Intercalation}},\ }\href
  {https://doi.org/10.1103/PhysRevLett.103.246804} {\bibfield  {journal}
  {\bibinfo  {journal} {Phys. Rev. Lett.}\ }\textbf {\bibinfo {volume} {103}},\
  \bibinfo {pages} {246804} (\bibinfo {year} {2009})},\ \Eprint
  {https://arxiv.org/abs/https://journals.aps.org/prl/abstract/10.1103/PhysRevLett.103.246804}
  {https://journals.aps.org/prl/abstract/10.1103/PhysRevLett.103.246804}
  \BibitemShut {NoStop}%
\bibitem [{\citenamefont {Gierz}\ \emph {et~al.}(2010)\citenamefont {Gierz},
  \citenamefont {Suzuki}, \citenamefont {Weitz}, \citenamefont {Lee},
  \citenamefont {Krauss}, \citenamefont {Riedl}, \citenamefont {Starke},
  \citenamefont {H\"ochst}, \citenamefont {Smet}, \citenamefont {Ast},\ and\
  \citenamefont {Kern}}]{Gierz2010}%
  \BibitemOpen
  \bibfield  {author} {\bibinfo {author} {\bibfnamefont {I.}~\bibnamefont
  {Gierz}}, \bibinfo {author} {\bibfnamefont {T.}~\bibnamefont {Suzuki}},
  \bibinfo {author} {\bibfnamefont {R.~T.}\ \bibnamefont {Weitz}}, \bibinfo
  {author} {\bibfnamefont {D.~S.}\ \bibnamefont {Lee}}, \bibinfo {author}
  {\bibfnamefont {B.}~\bibnamefont {Krauss}}, \bibinfo {author} {\bibfnamefont
  {C.}~\bibnamefont {Riedl}}, \bibinfo {author} {\bibfnamefont
  {U.}~\bibnamefont {Starke}}, \bibinfo {author} {\bibfnamefont
  {H.}~\bibnamefont {H\"ochst}}, \bibinfo {author} {\bibfnamefont {J.~H.}\
  \bibnamefont {Smet}}, \bibinfo {author} {\bibfnamefont {C.~R.}\ \bibnamefont
  {Ast}},\ and\ \bibinfo {author} {\bibfnamefont {K.}~\bibnamefont {Kern}},\
  }\bibfield  {title} {\bibinfo {title} {{Electronic decoupling of an epitaxial
  graphene monolayer by gold intercalation}},\ }\href
  {https://doi.org/10.1103/PhysRevB.81.235408} {\bibfield  {journal} {\bibinfo
  {journal} {Phys. Rev. B}\ }\textbf {\bibinfo {volume} {81}},\ \bibinfo
  {pages} {235408} (\bibinfo {year} {2010})},\ \Eprint
  {https://arxiv.org/abs/https://journals.aps.org/prb/abstract/10.1103/PhysRevB.81.235408}
  {https://journals.aps.org/prb/abstract/10.1103/PhysRevB.81.235408}
  \BibitemShut {NoStop}%
\bibitem [{\citenamefont {Briggs}\ \emph {et~al.}(2020)\citenamefont {Briggs},
  \citenamefont {Bersch}, \citenamefont {Wang}, \citenamefont {Jiang},
  \citenamefont {Koch}, \citenamefont {Nayir}, \citenamefont {Wang},
  \citenamefont {Kolmer}, \citenamefont {Ko}, \citenamefont {Duran},
  \citenamefont {Subramanian}, \citenamefont {Dong}, \citenamefont
  {Shallenberger}, \citenamefont {Fu}, \citenamefont {Zou}, \citenamefont
  {Chuang}, \citenamefont {Gai}, \citenamefont {Li}, \citenamefont {Bostwick},
  \citenamefont {Jozwiak}, \citenamefont {Chang}, \citenamefont {Rotenberg},
  \citenamefont {Zhu}, \citenamefont {van Duin}, \citenamefont {Crespi},\ and\
  \citenamefont {Robinson}}]{Briggs2020}%
  \BibitemOpen
  \bibfield  {author} {\bibinfo {author} {\bibfnamefont {N.}~\bibnamefont
  {Briggs}}, \bibinfo {author} {\bibfnamefont {B.}~\bibnamefont {Bersch}},
  \bibinfo {author} {\bibfnamefont {Y.}~\bibnamefont {Wang}}, \bibinfo {author}
  {\bibfnamefont {J.}~\bibnamefont {Jiang}}, \bibinfo {author} {\bibfnamefont
  {R.~J.}\ \bibnamefont {Koch}}, \bibinfo {author} {\bibfnamefont
  {N.}~\bibnamefont {Nayir}}, \bibinfo {author} {\bibfnamefont
  {K.}~\bibnamefont {Wang}}, \bibinfo {author} {\bibfnamefont {M.}~\bibnamefont
  {Kolmer}}, \bibinfo {author} {\bibfnamefont {W.}~\bibnamefont {Ko}}, \bibinfo
  {author} {\bibfnamefont {A.~D. L.~F.}\ \bibnamefont {Duran}}, \bibinfo
  {author} {\bibfnamefont {S.}~\bibnamefont {Subramanian}}, \bibinfo {author}
  {\bibfnamefont {C.}~\bibnamefont {Dong}}, \bibinfo {author} {\bibfnamefont
  {J.}~\bibnamefont {Shallenberger}}, \bibinfo {author} {\bibfnamefont
  {M.}~\bibnamefont {Fu}}, \bibinfo {author} {\bibfnamefont {Q.}~\bibnamefont
  {Zou}}, \bibinfo {author} {\bibfnamefont {Y.-W.}\ \bibnamefont {Chuang}},
  \bibinfo {author} {\bibfnamefont {Z.}~\bibnamefont {Gai}}, \bibinfo {author}
  {\bibfnamefont {A.-P.}\ \bibnamefont {Li}}, \bibinfo {author} {\bibfnamefont
  {A.}~\bibnamefont {Bostwick}}, \bibinfo {author} {\bibfnamefont
  {C.}~\bibnamefont {Jozwiak}}, \bibinfo {author} {\bibfnamefont {C.-Z.}\
  \bibnamefont {Chang}}, \bibinfo {author} {\bibfnamefont {E.}~\bibnamefont
  {Rotenberg}}, \bibinfo {author} {\bibfnamefont {J.}~\bibnamefont {Zhu}},
  \bibinfo {author} {\bibfnamefont {A.~C.~T.}\ \bibnamefont {van Duin}},
  \bibinfo {author} {\bibfnamefont {V.}~\bibnamefont {Crespi}},\ and\ \bibinfo
  {author} {\bibfnamefont {J.~A.}\ \bibnamefont {Robinson}},\ }\bibfield
  {title} {\bibinfo {title} {{Atomically thin half-van der Waals metals enabled
  by confinement heteroepitaxy}},\ }\href@noop {} {\bibfield  {journal}
  {\bibinfo  {journal} {Nature Materials}\ }\textbf {\bibinfo {volume} {19}},\
  \bibinfo {pages} {637} (\bibinfo {year} {2020})},\ \Eprint
  {https://arxiv.org/abs/https://www.nature.com/articles/s41563-020-0631-x}
  {https://www.nature.com/articles/s41563-020-0631-x} \BibitemShut {NoStop}%
\bibitem [{\citenamefont {Kim}\ \emph {et~al.}(2016)\citenamefont {Kim},
  \citenamefont {Dugerjav}, \citenamefont {Lkhagvasuren},\ and\ \citenamefont
  {Seo}}]{Kim2016}%
  \BibitemOpen
  \bibfield  {author} {\bibinfo {author} {\bibfnamefont {H.}~\bibnamefont
  {Kim}}, \bibinfo {author} {\bibfnamefont {O.}~\bibnamefont {Dugerjav}},
  \bibinfo {author} {\bibfnamefont {A.}~\bibnamefont {Lkhagvasuren}},\ and\
  \bibinfo {author} {\bibfnamefont {J.~M.}\ \bibnamefont {Seo}},\ }\bibfield
  {title} {\bibinfo {title} {{Charge neutrality of quasi-free-standing
  monolayer graphene induced by the intercalated Sn layer}},\ }\href
  {https://doi.org/10.1088/0022-3727/49/13/135307} {\bibfield  {journal}
  {\bibinfo  {journal} {Journal of Physics D: Applied Physics}\ }\textbf
  {\bibinfo {volume} {49}},\ \bibinfo {pages} {135307} (\bibinfo {year}
  {2016})},\ \Eprint
  {https://arxiv.org/abs/https://iopscience.iop.org/article/10.1088/0022-3727/49/13/135307}
  {https://iopscience.iop.org/article/10.1088/0022-3727/49/13/135307}
  \BibitemShut {NoStop}%
\bibitem [{\citenamefont {Hayashi}\ \emph {et~al.}(2017)\citenamefont
  {Hayashi}, \citenamefont {Visikovskiy}, \citenamefont {Kajiwara},
  \citenamefont {Iimori}, \citenamefont {Shirasawa}, \citenamefont {Nakastuji},
  \citenamefont {Miyamachi}, \citenamefont {Nakashima}, \citenamefont {Yaji},
  \citenamefont {Mase}, \citenamefont {Komori},\ and\ \citenamefont
  {Tanaka}}]{Hayashi2018}%
  \BibitemOpen
  \bibfield  {author} {\bibinfo {author} {\bibfnamefont {S.}~\bibnamefont
  {Hayashi}}, \bibinfo {author} {\bibfnamefont {A.}~\bibnamefont
  {Visikovskiy}}, \bibinfo {author} {\bibfnamefont {T.}~\bibnamefont
  {Kajiwara}}, \bibinfo {author} {\bibfnamefont {T.}~\bibnamefont {Iimori}},
  \bibinfo {author} {\bibfnamefont {T.}~\bibnamefont {Shirasawa}}, \bibinfo
  {author} {\bibfnamefont {K.}~\bibnamefont {Nakastuji}}, \bibinfo {author}
  {\bibfnamefont {T.}~\bibnamefont {Miyamachi}}, \bibinfo {author}
  {\bibfnamefont {S.}~\bibnamefont {Nakashima}}, \bibinfo {author}
  {\bibfnamefont {K.}~\bibnamefont {Yaji}}, \bibinfo {author} {\bibfnamefont
  {K.}~\bibnamefont {Mase}}, \bibinfo {author} {\bibfnamefont {F.}~\bibnamefont
  {Komori}},\ and\ \bibinfo {author} {\bibfnamefont {S.}~\bibnamefont
  {Tanaka}},\ }\bibfield  {title} {\bibinfo {title} {{Triangular lattice atomic
  layer of Sn $(1\times1)$ at graphene/SiC(0001) interface}},\ }\href
  {https://doi.org/10.7567/APEX.11.015202} {\bibfield  {journal} {\bibinfo
  {journal} {Applied Physics Express}\ }\textbf {\bibinfo {volume} {11}},\
  \bibinfo {pages} {015202} (\bibinfo {year} {2017})},\ \Eprint
  {https://arxiv.org/abs/https://iopscience.iop.org/article/10.7567/APEX.11.015202}
  {https://iopscience.iop.org/article/10.7567/APEX.11.015202} \BibitemShut
  {NoStop}%
\bibitem [{\citenamefont {Visikovskiy}\ \emph {et~al.}(2018)\citenamefont
  {Visikovskiy}, \citenamefont {Hayashi}, \citenamefont {Kajiwara},
  \citenamefont {Komori}, \citenamefont {Yaji},\ and\ \citenamefont
  {Tanaka}}]{Visikovskiy2018}%
  \BibitemOpen
  \bibfield  {author} {\bibinfo {author} {\bibfnamefont {A.}~\bibnamefont
  {Visikovskiy}}, \bibinfo {author} {\bibfnamefont {S.}~\bibnamefont
  {Hayashi}}, \bibinfo {author} {\bibfnamefont {T.}~\bibnamefont {Kajiwara}},
  \bibinfo {author} {\bibfnamefont {F.}~\bibnamefont {Komori}}, \bibinfo
  {author} {\bibfnamefont {K.}~\bibnamefont {Yaji}},\ and\ \bibinfo {author}
  {\bibfnamefont {S.}~\bibnamefont {Tanaka}},\ }\bibfield  {title} {\bibinfo
  {title} {{Computational study of heavy group IV elements (Ge, Sn, Pb)
  triangular lattice atomic layers on SiC(0001) surface}},\ }\href@noop {}
  {\bibfield  {journal} {\bibinfo  {journal} {arXiv:1809.00829}\ } (\bibinfo
  {year} {2018})},\ \Eprint
  {https://arxiv.org/abs/https://arxiv.org/abs/1809.00829}
  {https://arxiv.org/abs/1809.00829} \BibitemShut {NoStop}%
\bibitem [{\citenamefont {Yaji}\ \emph {et~al.}(2019)\citenamefont {Yaji},
  \citenamefont {Visikovskiy}, \citenamefont {Iimori}, \citenamefont {Kuroda},
  \citenamefont {Hayashi}, \citenamefont {Kajiwara}, \citenamefont {Tanaka},
  \citenamefont {Komori},\ and\ \citenamefont {Shin}}]{Yaji2019}%
  \BibitemOpen
  \bibfield  {author} {\bibinfo {author} {\bibfnamefont {K.}~\bibnamefont
  {Yaji}}, \bibinfo {author} {\bibfnamefont {A.}~\bibnamefont {Visikovskiy}},
  \bibinfo {author} {\bibfnamefont {T.}~\bibnamefont {Iimori}}, \bibinfo
  {author} {\bibfnamefont {K.}~\bibnamefont {Kuroda}}, \bibinfo {author}
  {\bibfnamefont {S.}~\bibnamefont {Hayashi}}, \bibinfo {author} {\bibfnamefont
  {T.}~\bibnamefont {Kajiwara}}, \bibinfo {author} {\bibfnamefont
  {S.}~\bibnamefont {Tanaka}}, \bibinfo {author} {\bibfnamefont
  {F.}~\bibnamefont {Komori}},\ and\ \bibinfo {author} {\bibfnamefont
  {S.}~\bibnamefont {Shin}},\ }\bibfield  {title} {\bibinfo {title}
  {{Coexistence of Two Types of Spin Splitting Originating from Different
  Symmetries}},\ }\href {https://doi.org/10.1103/PhysRevLett.122.126403}
  {\bibfield  {journal} {\bibinfo  {journal} {Phys. Rev. Lett.}\ }\textbf
  {\bibinfo {volume} {122}},\ \bibinfo {pages} {126403} (\bibinfo {year}
  {2019})},\ \Eprint
  {https://arxiv.org/abs/https://journals.aps.org/prl/abstract/10.1103/PhysRevLett.122.126403}
  {https://journals.aps.org/prl/abstract/10.1103/PhysRevLett.122.126403}
  \BibitemShut {NoStop}%
\bibitem [{\citenamefont {Rosenzweig}\ and\ \citenamefont
  {Starke}(2020)}]{Rosenzweig2020}%
  \BibitemOpen
  \bibfield  {author} {\bibinfo {author} {\bibfnamefont {P.}~\bibnamefont
  {Rosenzweig}}\ and\ \bibinfo {author} {\bibfnamefont {U.}~\bibnamefont
  {Starke}},\ }\bibfield  {title} {\bibinfo {title} {Large-area synthesis of a
  semiconducting silver monolayer via intercalation of epitaxial graphene},\
  }\href {https://doi.org/10.1103/PhysRevB.101.201407} {\bibfield  {journal}
  {\bibinfo  {journal} {Phys. Rev. B}\ }\textbf {\bibinfo {volume} {101}},\
  \bibinfo {pages} {201407} (\bibinfo {year} {2020})}\BibitemShut {NoStop}%
\bibitem [{\citenamefont {Forti}\ \emph {et~al.}(2020)\citenamefont {Forti},
  \citenamefont {Link}, \citenamefont {St\"ohr}, \citenamefont {Niu},
  \citenamefont {Zakharov}, \citenamefont {Coletti},\ and\ \citenamefont
  {Starke}}]{Forti2020}%
  \BibitemOpen
  \bibfield  {author} {\bibinfo {author} {\bibfnamefont {S.}~\bibnamefont
  {Forti}}, \bibinfo {author} {\bibfnamefont {S.}~\bibnamefont {Link}},
  \bibinfo {author} {\bibfnamefont {A.}~\bibnamefont {St\"ohr}}, \bibinfo
  {author} {\bibfnamefont {Y.}~\bibnamefont {Niu}}, \bibinfo {author}
  {\bibfnamefont {A.~A.}\ \bibnamefont {Zakharov}}, \bibinfo {author}
  {\bibfnamefont {C.}~\bibnamefont {Coletti}},\ and\ \bibinfo {author}
  {\bibfnamefont {U.}~\bibnamefont {Starke}},\ }\bibfield  {title} {\bibinfo
  {title} {Semiconductor to metal transition in two-dimensional gold and its
  van der waals heterostack with graphene},\ }\href
  {https://doi.org/10.1038/s41467-020-15683-1} {\bibfield  {journal} {\bibinfo
  {journal} {Nature Communications}\ }\textbf {\bibinfo {volume} {11}},\
  \bibinfo {pages} {2236} (\bibinfo {year} {2020})}\BibitemShut {NoStop}%
\bibitem [{\citenamefont {Glass}\ \emph {et~al.}(2015)\citenamefont {Glass},
  \citenamefont {Li}, \citenamefont {Adler}, \citenamefont {Aulbach},
  \citenamefont {Fleszar}, \citenamefont {Thomale}, \citenamefont {Hanke},
  \citenamefont {Claessen},\ and\ \citenamefont {Sch\"afer}}]{Glass2015}%
  \BibitemOpen
  \bibfield  {author} {\bibinfo {author} {\bibfnamefont {S.}~\bibnamefont
  {Glass}}, \bibinfo {author} {\bibfnamefont {G.}~\bibnamefont {Li}}, \bibinfo
  {author} {\bibfnamefont {F.}~\bibnamefont {Adler}}, \bibinfo {author}
  {\bibfnamefont {J.}~\bibnamefont {Aulbach}}, \bibinfo {author} {\bibfnamefont
  {A.}~\bibnamefont {Fleszar}}, \bibinfo {author} {\bibfnamefont
  {R.}~\bibnamefont {Thomale}}, \bibinfo {author} {\bibfnamefont
  {W.}~\bibnamefont {Hanke}}, \bibinfo {author} {\bibfnamefont
  {R.}~\bibnamefont {Claessen}},\ and\ \bibinfo {author} {\bibfnamefont
  {J.}~\bibnamefont {Sch\"afer}},\ }\bibfield  {title} {\bibinfo {title}
  {Triangular spin-orbit-coupled lattice with strong coulomb correlations: Sn
  atoms on a sic(0001) substrate},\ }\href
  {https://doi.org/10.1103/PhysRevLett.114.247602} {\bibfield  {journal}
  {\bibinfo  {journal} {Phys. Rev. Lett.}\ }\textbf {\bibinfo {volume} {114}},\
  \bibinfo {pages} {247602} (\bibinfo {year} {2015})}\BibitemShut {NoStop}%
\bibitem [{\citenamefont {Mamiyev}\ and\ \citenamefont
  {Tegenkamp}(2024)}]{Mamiyev2024}%
  \BibitemOpen
  \bibfield  {author} {\bibinfo {author} {\bibfnamefont {Z.}~\bibnamefont
  {Mamiyev}}\ and\ \bibinfo {author} {\bibfnamefont {C.}~\bibnamefont
  {Tegenkamp}},\ }\bibfield  {title} {\bibinfo {title} {Exploring
  graphene-substrate interactions: plasmonic excitation in sn-intercalated
  epitaxial graphene},\ }\href {https://doi.org/10.1088/2053-1583/ad1a70}
  {\bibfield  {journal} {\bibinfo  {journal} {2D Materials}\ }\textbf {\bibinfo
  {volume} {11}},\ \bibinfo {pages} {025013} (\bibinfo {year}
  {2024})}\BibitemShut {NoStop}%
\bibitem [{\citenamefont {Emtsev}\ \emph {et~al.}(2009)\citenamefont {Emtsev},
  \citenamefont {Bostwick}, \citenamefont {Horn}, \citenamefont {Jobst},
  \citenamefont {Kellogg}, \citenamefont {Ley}, \citenamefont {McChesney},
  \citenamefont {Ohta}, \citenamefont {Reshanov}, \citenamefont {R\"ohrl},
  \citenamefont {Rotenberg}, \citenamefont {Schmid}, \citenamefont {Waldmann},
  \citenamefont {Weber},\ and\ \citenamefont {Seyller}}]{Emtsev2009}%
  \BibitemOpen
  \bibfield  {author} {\bibinfo {author} {\bibfnamefont {K.~V.}\ \bibnamefont
  {Emtsev}}, \bibinfo {author} {\bibfnamefont {A.}~\bibnamefont {Bostwick}},
  \bibinfo {author} {\bibfnamefont {K.}~\bibnamefont {Horn}}, \bibinfo {author}
  {\bibfnamefont {J.}~\bibnamefont {Jobst}}, \bibinfo {author} {\bibfnamefont
  {G.~L.}\ \bibnamefont {Kellogg}}, \bibinfo {author} {\bibfnamefont
  {L.}~\bibnamefont {Ley}}, \bibinfo {author} {\bibfnamefont {J.~L.}\
  \bibnamefont {McChesney}}, \bibinfo {author} {\bibfnamefont {T.}~\bibnamefont
  {Ohta}}, \bibinfo {author} {\bibfnamefont {S.~A.}\ \bibnamefont {Reshanov}},
  \bibinfo {author} {\bibfnamefont {J.}~\bibnamefont {R\"ohrl}}, \bibinfo
  {author} {\bibfnamefont {E.}~\bibnamefont {Rotenberg}}, \bibinfo {author}
  {\bibfnamefont {A.~K.}\ \bibnamefont {Schmid}}, \bibinfo {author}
  {\bibfnamefont {D.}~\bibnamefont {Waldmann}}, \bibinfo {author}
  {\bibfnamefont {H.~B.}\ \bibnamefont {Weber}},\ and\ \bibinfo {author}
  {\bibfnamefont {T.}~\bibnamefont {Seyller}},\ }\bibfield  {title} {\bibinfo
  {title} {{Towards wafer-size graphene layers by atmospheric pressure
  graphitization of silicon carbide}},\ }\href@noop {} {\bibfield  {journal}
  {\bibinfo  {journal} {Nature Materials}\ }\textbf {\bibinfo {volume} {8}},\
  \bibinfo {pages} {203} (\bibinfo {year} {2009})},\ \Eprint
  {https://arxiv.org/abs/https://www.nature.com/articles/nmat2382}
  {https://www.nature.com/articles/nmat2382} \BibitemShut {NoStop}%
\bibitem [{\citenamefont {Varykhalov}(2018)}]{onesquare}%
  \BibitemOpen
  \bibfield  {author} {\bibinfo {author} {\bibfnamefont {A.}~\bibnamefont
  {Varykhalov}},\ }\bibfield  {title} {\bibinfo {title} {{$1^2$-ARPES: The
  ultra-high-resolution photoemission station at the U112-PGM-2a-$1^2$ beamline
  at BESSY II}},\ }\href@noop {} {\bibfield  {journal} {\bibinfo  {journal}
  {Journal of large-scale research facilities}\ }\textbf {\bibinfo {volume}
  {4}},\ \bibinfo {pages} {A128} (\bibinfo {year} {2018})},\ \Eprint
  {https://arxiv.org/abs/https://jlsrf.org/index.php/lsf/article/view/99}
  {https://jlsrf.org/index.php/lsf/article/view/99} \BibitemShut {NoStop}%
\bibitem [{\citenamefont {Kresse}\ and\ \citenamefont
  {Hafner}(1993)}]{Kresse1993}%
  \BibitemOpen
  \bibfield  {author} {\bibinfo {author} {\bibfnamefont {G.}~\bibnamefont
  {Kresse}}\ and\ \bibinfo {author} {\bibfnamefont {J.}~\bibnamefont
  {Hafner}},\ }\bibfield  {title} {\bibinfo {title} {{Ab initio molecular
  dynamics for liquid metals}},\ }\href
  {https://doi.org/10.1103/PhysRevB.47.558} {\bibfield  {journal} {\bibinfo
  {journal} {Phys. Rev. B}\ }\textbf {\bibinfo {volume} {47}},\ \bibinfo
  {pages} {558} (\bibinfo {year} {1993})},\ \Eprint
  {https://arxiv.org/abs/https://journals.aps.org/prb/abstract/10.1103/PhysRevB.47.558}
  {https://journals.aps.org/prb/abstract/10.1103/PhysRevB.47.558} \BibitemShut
  {NoStop}%
\bibitem [{\citenamefont {Kresse}\ and\ \citenamefont
  {Furthmüller}(1996)}]{Kresse1996}%
  \BibitemOpen
  \bibfield  {author} {\bibinfo {author} {\bibfnamefont {G.}~\bibnamefont
  {Kresse}}\ and\ \bibinfo {author} {\bibfnamefont {J.}~\bibnamefont
  {Furthmüller}},\ }\bibfield  {title} {\bibinfo {title} {{Efficiency of
  ab-initio total energy calculations for metals and semiconductors using a
  plane-wave basis set}},\ }\href
  {https://doi.org/https://doi.org/10.1016/0927-0256(96)00008-0} {\bibfield
  {journal} {\bibinfo  {journal} {Computational Materials Science}\ }\textbf
  {\bibinfo {volume} {6}},\ \bibinfo {pages} {15} (\bibinfo {year} {1996})},\
  \Eprint
  {https://arxiv.org/abs/https://www.sciencedirect.com/science/article/pii/0927025696000080}
  {https://www.sciencedirect.com/science/article/pii/0927025696000080}
  \BibitemShut {NoStop}%
\bibitem [{\citenamefont {Kresse}\ and\ \citenamefont
  {Furthm\"uller}(1996)}]{Kresse1996b}%
  \BibitemOpen
  \bibfield  {author} {\bibinfo {author} {\bibfnamefont {G.}~\bibnamefont
  {Kresse}}\ and\ \bibinfo {author} {\bibfnamefont {J.}~\bibnamefont
  {Furthm\"uller}},\ }\bibfield  {title} {\bibinfo {title} {{Efficient
  iterative schemes for ab initio total-energy calculations using a plane-wave
  basis set}},\ }\href {https://doi.org/10.1103/PhysRevB.54.11169} {\bibfield
  {journal} {\bibinfo  {journal} {Phys. Rev. B}\ }\textbf {\bibinfo {volume}
  {54}},\ \bibinfo {pages} {11169} (\bibinfo {year} {1996})},\ \Eprint
  {https://arxiv.org/abs/https://journals.aps.org/prb/abstract/10.1103/PhysRevB.54.11169}
  {https://journals.aps.org/prb/abstract/10.1103/PhysRevB.54.11169}
  \BibitemShut {NoStop}%
\bibitem [{\citenamefont {Perdew}\ \emph {et~al.}(1996)\citenamefont {Perdew},
  \citenamefont {Burke},\ and\ \citenamefont {Ernzerhof}}]{Perdew1996}%
  \BibitemOpen
  \bibfield  {author} {\bibinfo {author} {\bibfnamefont {J.~P.}\ \bibnamefont
  {Perdew}}, \bibinfo {author} {\bibfnamefont {K.}~\bibnamefont {Burke}},\ and\
  \bibinfo {author} {\bibfnamefont {M.}~\bibnamefont {Ernzerhof}},\ }\bibfield
  {title} {\bibinfo {title} {{Generalized Gradient Approximation Made
  Simple}},\ }\href {https://doi.org/10.1103/PhysRevLett.77.3865} {\bibfield
  {journal} {\bibinfo  {journal} {Phys. Rev. Lett.}\ }\textbf {\bibinfo
  {volume} {77}},\ \bibinfo {pages} {3865} (\bibinfo {year} {1996})},\ \Eprint
  {https://arxiv.org/abs/https://journals.aps.org/prl/abstract/10.1103/PhysRevLett.77.3865}
  {https://journals.aps.org/prl/abstract/10.1103/PhysRevLett.77.3865}
  \BibitemShut {NoStop}%
\bibitem [{\citenamefont {Bl\"ochl}(1994)}]{Bloechl1994}%
  \BibitemOpen
  \bibfield  {author} {\bibinfo {author} {\bibfnamefont {P.~E.}\ \bibnamefont
  {Bl\"ochl}},\ }\bibfield  {title} {\bibinfo {title} {{Projector
  augmented-wave method}},\ }\href {https://doi.org/10.1103/PhysRevB.50.17953}
  {\bibfield  {journal} {\bibinfo  {journal} {Phys. Rev. B}\ }\textbf {\bibinfo
  {volume} {50}},\ \bibinfo {pages} {17953} (\bibinfo {year} {1994})},\ \Eprint
  {https://arxiv.org/abs/https://journals.aps.org/prb/abstract/10.1103/PhysRevB.50.17953}
  {https://journals.aps.org/prb/abstract/10.1103/PhysRevB.50.17953}
  \BibitemShut {NoStop}%
\bibitem [{\citenamefont {Kresse}\ and\ \citenamefont
  {Joubert}(1999)}]{Kresse1999}%
  \BibitemOpen
  \bibfield  {author} {\bibinfo {author} {\bibfnamefont {G.}~\bibnamefont
  {Kresse}}\ and\ \bibinfo {author} {\bibfnamefont {D.}~\bibnamefont
  {Joubert}},\ }\bibfield  {title} {\bibinfo {title} {{From ultrasoft
  pseudopotentials to the projector augmented-wave method}},\ }\href
  {https://doi.org/10.1103/PhysRevB.59.1758} {\bibfield  {journal} {\bibinfo
  {journal} {Phys. Rev. B}\ }\textbf {\bibinfo {volume} {59}},\ \bibinfo
  {pages} {1758} (\bibinfo {year} {1999})},\ \Eprint
  {https://arxiv.org/abs/https://journals.aps.org/prb/abstract/10.1103/PhysRevB.59.1758}
  {https://journals.aps.org/prb/abstract/10.1103/PhysRevB.59.1758} \BibitemShut
  {NoStop}%
\bibitem [{\citenamefont {Mattausch}\ and\ \citenamefont
  {Pankratov}(2007)}]{Mattausch2007}%
  \BibitemOpen
  \bibfield  {author} {\bibinfo {author} {\bibfnamefont {A.}~\bibnamefont
  {Mattausch}}\ and\ \bibinfo {author} {\bibfnamefont {O.}~\bibnamefont
  {Pankratov}},\ }\bibfield  {title} {\bibinfo {title} {{Ab Initio Study of
  Graphene on SiC}},\ }\href {https://doi.org/10.1103/PhysRevLett.99.076802}
  {\bibfield  {journal} {\bibinfo  {journal} {Phys. Rev. Lett.}\ }\textbf
  {\bibinfo {volume} {99}},\ \bibinfo {pages} {076802} (\bibinfo {year}
  {2007})},\ \Eprint
  {https://arxiv.org/abs/https://journals.aps.org/prl/abstract/10.1103/PhysRevLett.99.076802}
  {https://journals.aps.org/prl/abstract/10.1103/PhysRevLett.99.076802}
  \BibitemShut {NoStop}%
\bibitem [{url()}]{url}%
  \BibitemOpen
  \href {https://dx.doi.org/10.17172/NOMAD/2024.04.06-1} {\ }\Eprint
  {https://arxiv.org/abs/https://dx.doi.org/10.17172/NOMAD/2024.04.06-1}
  {https://dx.doi.org/10.17172/NOMAD/2024.04.06-1} \BibitemShut {NoStop}%
\bibitem [{\citenamefont {Breusing}\ \emph {et~al.}(2009)\citenamefont
  {Breusing}, \citenamefont {Ropers},\ and\ \citenamefont
  {Elsaesser}}]{Breusing2009}%
  \BibitemOpen
  \bibfield  {author} {\bibinfo {author} {\bibfnamefont {M.}~\bibnamefont
  {Breusing}}, \bibinfo {author} {\bibfnamefont {C.}~\bibnamefont {Ropers}},\
  and\ \bibinfo {author} {\bibfnamefont {T.}~\bibnamefont {Elsaesser}},\
  }\bibfield  {title} {\bibinfo {title} {Ultrafast carrier dynamics in
  graphite},\ }\href {https://doi.org/10.1103/PhysRevLett.102.086809}
  {\bibfield  {journal} {\bibinfo  {journal} {Phys. Rev. Lett.}\ }\textbf
  {\bibinfo {volume} {102}},\ \bibinfo {pages} {086809} (\bibinfo {year}
  {2009})}\BibitemShut {NoStop}%
\bibitem [{\citenamefont {Kampfrath}\ \emph {et~al.}(2005)\citenamefont
  {Kampfrath}, \citenamefont {Perfetti}, \citenamefont {Schapper},
  \citenamefont {Frischkorn},\ and\ \citenamefont {Wolf}}]{Kampfrath2005}%
  \BibitemOpen
  \bibfield  {author} {\bibinfo {author} {\bibfnamefont {T.}~\bibnamefont
  {Kampfrath}}, \bibinfo {author} {\bibfnamefont {L.}~\bibnamefont {Perfetti}},
  \bibinfo {author} {\bibfnamefont {F.}~\bibnamefont {Schapper}}, \bibinfo
  {author} {\bibfnamefont {C.}~\bibnamefont {Frischkorn}},\ and\ \bibinfo
  {author} {\bibfnamefont {M.}~\bibnamefont {Wolf}},\ }\bibfield  {title}
  {\bibinfo {title} {Strongly coupled optical phonons in the ultrafast dynamics
  of the electronic energy and current relaxation in graphite},\ }\href
  {https://doi.org/10.1103/PhysRevLett.95.187403} {\bibfield  {journal}
  {\bibinfo  {journal} {Phys. Rev. Lett.}\ }\textbf {\bibinfo {volume} {95}},\
  \bibinfo {pages} {187403} (\bibinfo {year} {2005})}\BibitemShut {NoStop}%
\bibitem [{\citenamefont {Yan}\ \emph {et~al.}(2009)\citenamefont {Yan},
  \citenamefont {Song}, \citenamefont {Mak}, \citenamefont {Chatzakis},
  \citenamefont {Maultzsch},\ and\ \citenamefont {Heinz}}]{Yan2009}%
  \BibitemOpen
  \bibfield  {author} {\bibinfo {author} {\bibfnamefont {H.}~\bibnamefont
  {Yan}}, \bibinfo {author} {\bibfnamefont {D.}~\bibnamefont {Song}}, \bibinfo
  {author} {\bibfnamefont {K.~F.}\ \bibnamefont {Mak}}, \bibinfo {author}
  {\bibfnamefont {I.}~\bibnamefont {Chatzakis}}, \bibinfo {author}
  {\bibfnamefont {J.}~\bibnamefont {Maultzsch}},\ and\ \bibinfo {author}
  {\bibfnamefont {T.~F.}\ \bibnamefont {Heinz}},\ }\bibfield  {title} {\bibinfo
  {title} {Time-resolved raman spectroscopy of optical phonons in graphite:
  Phonon anharmonic coupling and anomalous stiffening},\ }\href
  {https://doi.org/10.1103/PhysRevB.80.121403} {\bibfield  {journal} {\bibinfo
  {journal} {Phys. Rev. B}\ }\textbf {\bibinfo {volume} {80}},\ \bibinfo
  {pages} {121403} (\bibinfo {year} {2009})}\BibitemShut {NoStop}%
\bibitem [{\citenamefont {Johannsen}\ \emph {et~al.}(2013)\citenamefont
  {Johannsen}, \citenamefont {Ulstrup}, \citenamefont {Cilento}, \citenamefont
  {Crepaldi}, \citenamefont {Zacchigna}, \citenamefont {Cacho}, \citenamefont
  {Turcu}, \citenamefont {Springate}, \citenamefont {Fromm}, \citenamefont
  {Raidel}, \citenamefont {Seyller}, \citenamefont {Parmigiani}, \citenamefont
  {Grioni},\ and\ \citenamefont {Hofmann}}]{Johannsen2013}%
  \BibitemOpen
  \bibfield  {author} {\bibinfo {author} {\bibfnamefont {J.~C.}\ \bibnamefont
  {Johannsen}}, \bibinfo {author} {\bibfnamefont {S.}~\bibnamefont {Ulstrup}},
  \bibinfo {author} {\bibfnamefont {F.}~\bibnamefont {Cilento}}, \bibinfo
  {author} {\bibfnamefont {A.}~\bibnamefont {Crepaldi}}, \bibinfo {author}
  {\bibfnamefont {M.}~\bibnamefont {Zacchigna}}, \bibinfo {author}
  {\bibfnamefont {C.}~\bibnamefont {Cacho}}, \bibinfo {author} {\bibfnamefont
  {I.~C.~E.}\ \bibnamefont {Turcu}}, \bibinfo {author} {\bibfnamefont
  {E.}~\bibnamefont {Springate}}, \bibinfo {author} {\bibfnamefont
  {F.}~\bibnamefont {Fromm}}, \bibinfo {author} {\bibfnamefont
  {C.}~\bibnamefont {Raidel}}, \bibinfo {author} {\bibfnamefont
  {T.}~\bibnamefont {Seyller}}, \bibinfo {author} {\bibfnamefont
  {F.}~\bibnamefont {Parmigiani}}, \bibinfo {author} {\bibfnamefont
  {M.}~\bibnamefont {Grioni}},\ and\ \bibinfo {author} {\bibfnamefont
  {P.}~\bibnamefont {Hofmann}},\ }\bibfield  {title} {\bibinfo {title} {Direct
  view of hot carrier dynamics in graphene},\ }\href
  {https://doi.org/10.1103/PhysRevLett.111.027403} {\bibfield  {journal}
  {\bibinfo  {journal} {Phys. Rev. Lett.}\ }\textbf {\bibinfo {volume} {111}},\
  \bibinfo {pages} {027403} (\bibinfo {year} {2013})}\BibitemShut {NoStop}%
\bibitem [{\citenamefont {Gierz}\ \emph {et~al.}(2014)\citenamefont {Gierz},
  \citenamefont {Link}, \citenamefont {Starke},\ and\ \citenamefont
  {Cavalleri}}]{GierzFD2014}%
  \BibitemOpen
  \bibfield  {author} {\bibinfo {author} {\bibfnamefont {I.}~\bibnamefont
  {Gierz}}, \bibinfo {author} {\bibfnamefont {S.}~\bibnamefont {Link}},
  \bibinfo {author} {\bibfnamefont {U.}~\bibnamefont {Starke}},\ and\ \bibinfo
  {author} {\bibfnamefont {A.}~\bibnamefont {Cavalleri}},\ }\bibfield  {title}
  {\bibinfo {title} {{Non-equilibrium Dirac carrier dynamics in graphene
  investigated with time- and angle-resolved photoemission spectroscopy}},\
  }\href {https://doi.org/10.1039/C4FD00020J} {\bibfield  {journal} {\bibinfo
  {journal} {Faraday Discuss.}\ }\textbf {\bibinfo {volume} {171}},\ \bibinfo
  {pages} {311} (\bibinfo {year} {2014})},\ \Eprint
  {https://arxiv.org/abs/https://pubs.rsc.org/en/content/articlelanding/2014/fd/c4fd00020j}
  {https://pubs.rsc.org/en/content/articlelanding/2014/fd/c4fd00020j}
  \BibitemShut {NoStop}%
\bibitem [{\citenamefont {Gierz}\ \emph {et~al.}(2015)\citenamefont {Gierz},
  \citenamefont {Mitrano}, \citenamefont {Petersen}, \citenamefont {Cacho},
  \citenamefont {Turcu}, \citenamefont {Springate}, \citenamefont {St\"ohr},
  \citenamefont {K\"ohler}, \citenamefont {Starke},\ and\ \citenamefont
  {Cavalleri}}]{GierzJPCM2015}%
  \BibitemOpen
  \bibfield  {author} {\bibinfo {author} {\bibfnamefont {I.}~\bibnamefont
  {Gierz}}, \bibinfo {author} {\bibfnamefont {M.}~\bibnamefont {Mitrano}},
  \bibinfo {author} {\bibfnamefont {J.~C.}\ \bibnamefont {Petersen}}, \bibinfo
  {author} {\bibfnamefont {C.}~\bibnamefont {Cacho}}, \bibinfo {author}
  {\bibfnamefont {I.~C.~E.}\ \bibnamefont {Turcu}}, \bibinfo {author}
  {\bibfnamefont {E.}~\bibnamefont {Springate}}, \bibinfo {author}
  {\bibfnamefont {A.}~\bibnamefont {St\"ohr}}, \bibinfo {author} {\bibfnamefont
  {A.}~\bibnamefont {K\"ohler}}, \bibinfo {author} {\bibfnamefont
  {U.}~\bibnamefont {Starke}},\ and\ \bibinfo {author} {\bibfnamefont
  {A.}~\bibnamefont {Cavalleri}},\ }\bibfield  {title} {\bibinfo {title}
  {{Population inversion in monolayer and bilayer graphene}},\ }\href
  {https://doi.org/10.1088/0953-8984/27/16/164204} {\bibfield  {journal}
  {\bibinfo  {journal} {Journal of Physics: Condensed Matter}\ }\textbf
  {\bibinfo {volume} {27}},\ \bibinfo {pages} {164204} (\bibinfo {year}
  {2015})},\ \Eprint
  {https://arxiv.org/abs/https://iopscience.iop.org/article/10.1088/0953-8984/27/16/164204}
  {https://iopscience.iop.org/article/10.1088/0953-8984/27/16/164204}
  \BibitemShut {NoStop}%
\bibitem [{\citenamefont {Johannsen}\ \emph {et~al.}(2015)\citenamefont
  {Johannsen}, \citenamefont {Ulstrup}, \citenamefont {Crepaldi}, \citenamefont
  {Cilento}, \citenamefont {Zacchigna}, \citenamefont {Miwa}, \citenamefont
  {Cacho}, \citenamefont {Chapman}, \citenamefont {Springate}, \citenamefont
  {Fromm}, \citenamefont {Raidel}, \citenamefont {Seyller}, \citenamefont
  {King}, \citenamefont {Parmigiani}, \citenamefont {Grioni},\ and\
  \citenamefont {Hofmann}}]{Johannsen2015}%
  \BibitemOpen
  \bibfield  {author} {\bibinfo {author} {\bibfnamefont {J.~C.}\ \bibnamefont
  {Johannsen}}, \bibinfo {author} {\bibfnamefont {S.}~\bibnamefont {Ulstrup}},
  \bibinfo {author} {\bibfnamefont {A.}~\bibnamefont {Crepaldi}}, \bibinfo
  {author} {\bibfnamefont {F.}~\bibnamefont {Cilento}}, \bibinfo {author}
  {\bibfnamefont {M.}~\bibnamefont {Zacchigna}}, \bibinfo {author}
  {\bibfnamefont {J.~A.}\ \bibnamefont {Miwa}}, \bibinfo {author}
  {\bibfnamefont {C.}~\bibnamefont {Cacho}}, \bibinfo {author} {\bibfnamefont
  {R.~T.}\ \bibnamefont {Chapman}}, \bibinfo {author} {\bibfnamefont
  {E.}~\bibnamefont {Springate}}, \bibinfo {author} {\bibfnamefont
  {F.}~\bibnamefont {Fromm}}, \bibinfo {author} {\bibfnamefont
  {C.}~\bibnamefont {Raidel}}, \bibinfo {author} {\bibfnamefont
  {T.}~\bibnamefont {Seyller}}, \bibinfo {author} {\bibfnamefont {P.~D.~C.}\
  \bibnamefont {King}}, \bibinfo {author} {\bibfnamefont {F.}~\bibnamefont
  {Parmigiani}}, \bibinfo {author} {\bibfnamefont {M.}~\bibnamefont {Grioni}},\
  and\ \bibinfo {author} {\bibfnamefont {P.}~\bibnamefont {Hofmann}},\
  }\bibfield  {title} {\bibinfo {title} {{Tunable Carrier Multiplication and
  Cooling in Graphene}},\ }\href {https://doi.org/10.1021/nl503614v} {\bibfield
   {journal} {\bibinfo  {journal} {Nano Letters}\ }\textbf {\bibinfo {volume}
  {15}},\ \bibinfo {pages} {326} (\bibinfo {year} {2015})},\ \Eprint
  {https://arxiv.org/abs/https://pubs.acs.org/doi/full/10.1021/nl503614v}
  {https://pubs.acs.org/doi/full/10.1021/nl503614v} \BibitemShut {NoStop}%
\bibitem [{\citenamefont {Ulstrup}\ \emph {et~al.}(2015)\citenamefont
  {Ulstrup}, \citenamefont {Johannsen}, \citenamefont {Crepaldi}, \citenamefont
  {Cilento}, \citenamefont {Zacchigna}, \citenamefont {Cacho}, \citenamefont
  {Chapman}, \citenamefont {Springate}, \citenamefont {Fromm}, \citenamefont
  {Raidel}, \citenamefont {Seyller}, \citenamefont {Parmigiani}, \citenamefont
  {Grioni},\ and\ \citenamefont {Hofmann}}]{Ulstrup2015}%
  \BibitemOpen
  \bibfield  {author} {\bibinfo {author} {\bibfnamefont {S.}~\bibnamefont
  {Ulstrup}}, \bibinfo {author} {\bibfnamefont {J.~C.}\ \bibnamefont
  {Johannsen}}, \bibinfo {author} {\bibfnamefont {A.}~\bibnamefont {Crepaldi}},
  \bibinfo {author} {\bibfnamefont {F.}~\bibnamefont {Cilento}}, \bibinfo
  {author} {\bibfnamefont {M.}~\bibnamefont {Zacchigna}}, \bibinfo {author}
  {\bibfnamefont {C.}~\bibnamefont {Cacho}}, \bibinfo {author} {\bibfnamefont
  {R.~T.}\ \bibnamefont {Chapman}}, \bibinfo {author} {\bibfnamefont
  {E.}~\bibnamefont {Springate}}, \bibinfo {author} {\bibfnamefont
  {F.}~\bibnamefont {Fromm}}, \bibinfo {author} {\bibfnamefont
  {C.}~\bibnamefont {Raidel}}, \bibinfo {author} {\bibfnamefont
  {T.}~\bibnamefont {Seyller}}, \bibinfo {author} {\bibfnamefont
  {F.}~\bibnamefont {Parmigiani}}, \bibinfo {author} {\bibfnamefont
  {M.}~\bibnamefont {Grioni}},\ and\ \bibinfo {author} {\bibfnamefont
  {P.}~\bibnamefont {Hofmann}},\ }\bibfield  {title} {\bibinfo {title}
  {{Ultrafast electron dynamics in epitaxial graphene investigated with time-
  and angle-resolved photoemission spectroscopy}},\ }\href
  {https://doi.org/10.1088/0953-8984/27/16/164206} {\bibfield  {journal}
  {\bibinfo  {journal} {Journal of Physics: Condensed Matter}\ }\textbf
  {\bibinfo {volume} {27}},\ \bibinfo {pages} {164206} (\bibinfo {year}
  {2015})},\ \Eprint
  {https://arxiv.org/abs/https://iopscience.iop.org/article/10.1088/0953-8984/27/16/164206}
  {https://iopscience.iop.org/article/10.1088/0953-8984/27/16/164206}
  \BibitemShut {NoStop}%
\bibitem [{\citenamefont {Majchrzak}\ \emph {et~al.}(2021)\citenamefont
  {Majchrzak}, \citenamefont {Volckaert}, \citenamefont {Čabo}, \citenamefont
  {Biswas}, \citenamefont {Bianchi}, \citenamefont {Mahatha}, \citenamefont
  {Dendzik}, \citenamefont {Andreatta}, \citenamefont {Grønborg},
  \citenamefont {Marković}, \citenamefont {Riley}, \citenamefont {Johannsen},
  \citenamefont {Lizzit}, \citenamefont {Bignardi}, \citenamefont {Lizzit},
  \citenamefont {Cacho}, \citenamefont {Alexander}, \citenamefont {Matselyukh},
  \citenamefont {Wyatt}, \citenamefont {Chapman}, \citenamefont {Springate},
  \citenamefont {Lauritsen}, \citenamefont {King}, \citenamefont {Sanders},
  \citenamefont {Miwa}, \citenamefont {Hofmann},\ and\ \citenamefont
  {Ulstrup}}]{Majchrzak2021}%
  \BibitemOpen
  \bibfield  {author} {\bibinfo {author} {\bibfnamefont {P.}~\bibnamefont
  {Majchrzak}}, \bibinfo {author} {\bibfnamefont {K.}~\bibnamefont
  {Volckaert}}, \bibinfo {author} {\bibfnamefont {A.~G.}\ \bibnamefont
  {Čabo}}, \bibinfo {author} {\bibfnamefont {D.}~\bibnamefont {Biswas}},
  \bibinfo {author} {\bibfnamefont {M.}~\bibnamefont {Bianchi}}, \bibinfo
  {author} {\bibfnamefont {S.~K.}\ \bibnamefont {Mahatha}}, \bibinfo {author}
  {\bibfnamefont {M.}~\bibnamefont {Dendzik}}, \bibinfo {author} {\bibfnamefont
  {F.}~\bibnamefont {Andreatta}}, \bibinfo {author} {\bibfnamefont {S.~S.}\
  \bibnamefont {Grønborg}}, \bibinfo {author} {\bibfnamefont {I.}~\bibnamefont
  {Marković}}, \bibinfo {author} {\bibfnamefont {J.~M.}\ \bibnamefont
  {Riley}}, \bibinfo {author} {\bibfnamefont {J.~C.}\ \bibnamefont
  {Johannsen}}, \bibinfo {author} {\bibfnamefont {D.}~\bibnamefont {Lizzit}},
  \bibinfo {author} {\bibfnamefont {L.}~\bibnamefont {Bignardi}}, \bibinfo
  {author} {\bibfnamefont {S.}~\bibnamefont {Lizzit}}, \bibinfo {author}
  {\bibfnamefont {C.}~\bibnamefont {Cacho}}, \bibinfo {author} {\bibfnamefont
  {O.}~\bibnamefont {Alexander}}, \bibinfo {author} {\bibfnamefont
  {D.}~\bibnamefont {Matselyukh}}, \bibinfo {author} {\bibfnamefont {A.~S.}\
  \bibnamefont {Wyatt}}, \bibinfo {author} {\bibfnamefont {R.~T.}\ \bibnamefont
  {Chapman}}, \bibinfo {author} {\bibfnamefont {E.}~\bibnamefont {Springate}},
  \bibinfo {author} {\bibfnamefont {J.~V.}\ \bibnamefont {Lauritsen}}, \bibinfo
  {author} {\bibfnamefont {P.~D.}\ \bibnamefont {King}}, \bibinfo {author}
  {\bibfnamefont {C.~E.}\ \bibnamefont {Sanders}}, \bibinfo {author}
  {\bibfnamefont {J.~A.}\ \bibnamefont {Miwa}}, \bibinfo {author}
  {\bibfnamefont {P.}~\bibnamefont {Hofmann}},\ and\ \bibinfo {author}
  {\bibfnamefont {S.}~\bibnamefont {Ulstrup}},\ }\bibfield  {title} {\bibinfo
  {title} {{Spectroscopic view of ultrafast charge carrier dynamics in single-
  and bilayer transition metal dichalcogenide semiconductors}},\ }\href
  {https://doi.org/https://doi.org/10.1016/j.elspec.2021.147093} {\bibfield
  {journal} {\bibinfo  {journal} {Journal of Electron Spectroscopy and Related
  Phenomena}\ }\textbf {\bibinfo {volume} {250}},\ \bibinfo {pages} {147093}
  (\bibinfo {year} {2021})},\ \Eprint
  {https://arxiv.org/abs/https://www.sciencedirect.com/science/article/pii/S0368204821000475}
  {https://www.sciencedirect.com/science/article/pii/S0368204821000475}
  \BibitemShut {NoStop}%
\bibitem [{\citenamefont {Hofmann}\ \emph {et~al.}(2023)\citenamefont
  {Hofmann}, \citenamefont {Weigl}, \citenamefont {Gradl}, \citenamefont
  {Mishra}, \citenamefont {Orlandini}, \citenamefont {Forti}, \citenamefont
  {Coletti}, \citenamefont {Latini}, \citenamefont {Xian}, \citenamefont
  {Rubio}, \citenamefont {Paredes}, \citenamefont {Causin}, \citenamefont
  {Brem}, \citenamefont {Malic},\ and\ \citenamefont {Gierz}}]{Hofmann2023}%
  \BibitemOpen
  \bibfield  {author} {\bibinfo {author} {\bibfnamefont {N.}~\bibnamefont
  {Hofmann}}, \bibinfo {author} {\bibfnamefont {L.}~\bibnamefont {Weigl}},
  \bibinfo {author} {\bibfnamefont {J.}~\bibnamefont {Gradl}}, \bibinfo
  {author} {\bibfnamefont {N.}~\bibnamefont {Mishra}}, \bibinfo {author}
  {\bibfnamefont {G.}~\bibnamefont {Orlandini}}, \bibinfo {author}
  {\bibfnamefont {S.}~\bibnamefont {Forti}}, \bibinfo {author} {\bibfnamefont
  {C.}~\bibnamefont {Coletti}}, \bibinfo {author} {\bibfnamefont
  {S.}~\bibnamefont {Latini}}, \bibinfo {author} {\bibfnamefont
  {L.}~\bibnamefont {Xian}}, \bibinfo {author} {\bibfnamefont {A.}~\bibnamefont
  {Rubio}}, \bibinfo {author} {\bibfnamefont {D.~P.}\ \bibnamefont {Paredes}},
  \bibinfo {author} {\bibfnamefont {R.~P.}\ \bibnamefont {Causin}}, \bibinfo
  {author} {\bibfnamefont {S.}~\bibnamefont {Brem}}, \bibinfo {author}
  {\bibfnamefont {E.}~\bibnamefont {Malic}},\ and\ \bibinfo {author}
  {\bibfnamefont {I.}~\bibnamefont {Gierz}},\ }\bibfield  {title} {\bibinfo
  {title} {{Link between interlayer hybridization and ultrafast charge transfer
  in WS$_2$-graphene heterostructures}},\ }\href
  {https://doi.org/10.1088/2053-1583/acdaab} {\bibfield  {journal} {\bibinfo
  {journal} {2D Materials}\ }\textbf {\bibinfo {volume} {10}},\ \bibinfo
  {pages} {035025} (\bibinfo {year} {2023})},\ \Eprint
  {https://arxiv.org/abs/https://iopscience.iop.org/article/10.1088/2053-1583/acdaab}
  {https://iopscience.iop.org/article/10.1088/2053-1583/acdaab} \BibitemShut
  {NoStop}%
\end{thebibliography}%

\pagebreak
	
	\end{document}